\documentclass{fairmeta}
\usepackage{amsmath}
\usepackage{enumerate} 
\usepackage{bm}
\usepackage{algorithm}
\usepackage{floatflt}
\usepackage{algpseudocode}
\usepackage{amsfonts}
\usepackage{amsthm}
\usepackage{newtxtt}
\usepackage{algorithm}
\usepackage{colortbl} 
\usepackage{cleveref}
\usepackage{diagbox} 
\usepackage[utf8]{inputenc}
\usepackage{textgreek}
\usepackage{colortbl}
\usepackage{nicematrix}
\usepackage{makecell}
\usepackage{float}
\usepackage{arydshln}
\usepackage[frozencache,cachedir=.]{minted}
\usepackage{caption}
\usepackage{subcaption}
\usepackage{tcolorbox}
\usepackage{amssymb}
\usepackage{xspace}
\usepackage{wrapfig}
\usepackage{adjustbox}
\usepackage{tabularx}
\usepackage{booktabs}
\usepackage{mathtools}
\usepackage{graphicx}
\usepackage{caption}
\usepackage{silence}
\makeatletter
\patchcmd{\wrong@fontshape}{\@gobbletwo}{}{}{}
\makeatother
\definecolor{upColor}{RGB}{17,138,21}
\definecolor{downColor}{RGB}{174,36,67}

\newtheorem{theorem}{Theorem}[]

\newtheorem{remark1}[theorem]{Remark}

\theoremstyle{plain}

\title{Bridging Visual and Wireless Sensing via a Unified Radiation Field for 3D Radio Map Construction}
\author[1]{Chaozheng Wen}
\author[1]{Jingwen Tong}
\author[1]{Zehong Lin}
\author[1]{Chenghong Bian}
\author[1]{Jun Zhang}
\affiliation[1]{Hong Kong University of Science and Technology}
\code{\url{https://github.com/wenchaozheng/URF-GS }}  % 示例，可替换
\metadata[Correspondence to]{Jun Zhang (\email{eejzhang@ust.hk})}

\begin{document}

\abstract{
The emerging applications of next-generation wireless networks demand high-fidelity environmental intelligence. 3D radio maps bridge physical environments and electromagnetic propagation for spectrum planning and environment-aware sensing. However, most existing methods treat visual and wireless data as independent modalities and fail to leverage shared electromagnetic propagation principles. To bridge this gap, we propose \textit{URF-GS}, a unified radio-optical radiation field framework based on 3D Gaussian splatting and inverse rendering for 3D radio map construction. By fusing cross-modal observations, our method recovers scene geometry and material properties to predict radio signals under arbitrary transceiver configurations without retraining. Experiments demonstrate up to a $24.7\%$ improvement in spatial spectrum accuracy and a $10\times$ increase in sample efficiency compared with NeRF-based methods. We further showcase URF-GS in Wi-Fi AP deployment and robot path planning tasks. This unified visual-wireless representation supports holistic radiation field modeling for future wireless communication systems.
}

\maketitle

\section{Introduction}
\label{sec1}

Wireless networks have evolved into a critical infrastructure for modern society, supporting a vast ecosystem of interconnected devices and applications. As next-generation systems strive to merge physical and virtual realms through immersive 3D experiences~\cite{wu2024embracing}, and support diverse scenarios through ubiquitous connectivity (e.g., low-altitude networks \cite{wu2025low}), the complexity of network design and optimization has escalated. To address these challenges, the concept of 3D radio maps has emerged as a pivotal solution. By providing a spatial representation of electromagnetic fields, the 3D radio map connect the physical environment with its embedded wireless communication channels~\cite{bi2019engineering}. Such maps are indispensable for spectrum-aware planning, proactive interference management~\cite{hu20233d}, and emerging applications such as low-altitude economy and indoor service robots~\cite{esrafilian20173d}. Furthermore, with the rise of integrated sensing and communication (ISAC) and wireless imaging, high-fidelity 3D radio maps are becoming the cornerstone for advanced tasks such as CSI-based imaging~\cite{bazzi2025isac}, hybrid radar fusion~\cite{chowdary2024hybrid}, and sensing-centric waveform optimization~\cite{bazzi2025designing}. Essentially, 3D radio maps serve as critical foundation for deploying intelligent wireless systems in complex physical environments.

However, it is highly non-trivial to construct high-fidelity 3D radio maps. Existing methodologies mainly rely on wireless channel modeling, which can be categorized into probabilistic ~\cite{sarkar2003survey}, deterministic ~\cite{he2018design}, and learning-based models~\cite{RadioUNet, AI_radiomap}. In particular, probabilistic approaches ~\cite{sarkar2003survey} overly simplify radio propagation via metrics such as the path loss exponent, while deterministic methods, e.g., ray-tracing~\cite{he2018design}, require precise computer-aided design models as well as accurate geometry and material information that are rarely accessible. Meanwhile, learning-based models ~\cite{RadioUNet} often act as black boxes, lacking interpretability and physical priors, and suffer from high computational complexity.

Despite these challenges, a new generation of physics-informed frameworks and foundation models has emerged to incorporate electromagnetic priors into radio map construction. Notable examples include RMDM~\cite{jia2025rmdm}, which utilizes physics-informed diffusion models, ReVeal~\cite{shahid2025reveal}, which employs physics-informed neural networks (PINNs), and PEFNet~\cite{jiang2024physics}, which solves electromagnetic integral equations. Additionally, the CKM foundation model~\cite{xiao2026scalable} explores a scalable collaborative framework for radio environmental awareness. Despite their progress, these methodologies largely formulate radio map construction as a grid-regression or image-restoration task. Such 2D/2.5D-centric paradigms lack an explicit 3D geometric representation of the environment, making them inadequate for capturing complex propagation variations across different altitudes or handling non-planar obstacles. Furthermore, many of these methods rely on iterative generative processes or intensive numerical solvers, which prohibit the real-time inference required for dynamic network optimization.

Recent advances in computer vision, particularly neural radiance fields (NeRF)~\cite{mildenhall2021nerf} and 3D Gaussian splatting (3D-GS)~\cite{kerbl3Dgaussians}, provide a promising avenue for accurate 3D radio map construction. These techniques approximate light transport via volumetric rendering, circumventing the need for computationally prohibitive full-wave electromagnetic solvers. While light and RF signals differ in their interaction mechanisms (e.g., diffraction and penetration), their shared wave nature allows visual reconstruction pipelines to provide a strong geometric prior that significantly simplifies the subsequent wireless channel modeling. For instance, NeRF\textsuperscript{2}~\cite{zhao2023nerf2} and WRF-GS~\cite{wen2025wrf, zhang2024rf} utilize NeRF and 3D-GS for 3D radio map construction based on radiation field reconstruction, respectively. Although RF-3DGS~\cite{zhang2024rf} improves construction accuracy by integrating visual and wireless information, the reconstructed radiation field is overfit to the fixed transmitter-receiver (Tx-Rx) configuration, and generalization across different Tx or Rx locations remains a major challenge. Moreover, existing approaches often fail to explicitly learn the underlying geometry and material properties, as well as radiation patterns, making robust adaptation to the Tx position and network configuration impossible. This substantially limits their capability to handle various tasks such as network planning and access point (AP) deployment~\cite{cao2025photon}, and results in a pronounced gap between simulation and real-world deployment.

\begin{figure}[!t]
\centering
\setlength{\fboxsep}{0pt}
\includegraphics[width=1\textwidth]{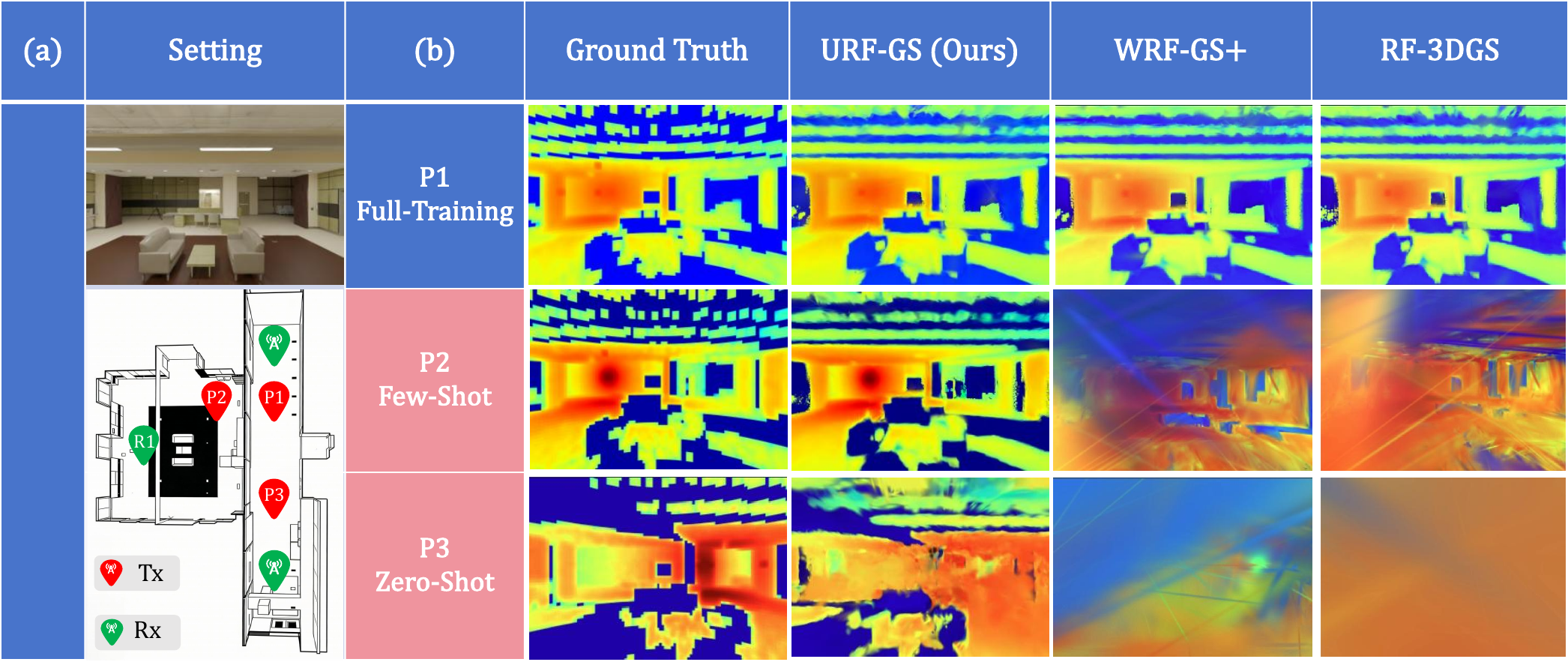}
\caption{\textbf{Comparison of constructed 3D radio maps in an indoor environment using URF-GS versus various baselines.} (a) \textbf{Environmental Setting}: The upper image presents the room viewed from the visual sensor (i.e., a camera). The bottom image displays the top-down layout of the indoor environment where the Rx can move arbitrarily within this space. Three Tx locations (P1–P3) are evaluated: a large number of samples (of different RX positions) are collected for P1, whereas P2 (10 samples) and P3 (zero-shot) have access to a limited number of samples. (b) \textbf{Constructed Radio Maps}: The 3D radio maps generated at Rx location R1 via different methods, corresponding to the specified Tx locations. URF-GS achieves the highest prediction accuracy among the baselines, closely matching the ground truth in all scenarios (i.e., across locations P1–P3).}\label{SynSS}
\end{figure}

In this article, we propose URF-GS, a \emph{Unified} radio-optical \emph{Radiation Field} representation framework for accurate and generalizable 3D radio map construction. Built upon 3D-GS and inverse rendering~\cite{liang2024gs}, URF-GS employs a unified representation to jointly characterize both the radio and optical fields using 3D Gaussian primitives. Unlike previous methods that are confined to fixed configurations, URF-GS leverages physics-informed inverse rendering to learn material properties and radiation patterns from multi-modal observations. This enables the system to perform spatial ray-tracing and generalize to different Tx-Rx configurations. As illustrated in Fig. \ref{SynSS}, URF-GS yields higher accuracy than baseline schemes, closely matches the ground truth (GT) across all locations and dataset sizes. These results underscore the strong location generalization capability of URF-GS under arbitrary Tx-Rx configurations. More quantitative results and case studies are presented in the next section.

\section{Results}\label{sec2}
We conduct extensive experiments to evaluate URF-GS.
We first implement it in a 3D environment to evaluate the 3D radio map construction accuracy and the generalization ability to different Tx-Rx configurations. 
The numerical results demonstrate that URF-GS attains a higher construction accuracy than baselines. Notably, it achieves an improvement of up to $\mathbf{24.7}\%$ in spatial spectrum prediction accuracy and more than $\mathbf{10\times}$ in sample efficiency for the construction of 3D radio maps compared to NeRF-based methods. Then, we test URF-GS in practical application scenarios, including Wi-Fi AP deployment and robot path planning. Experiment results show that URF-GS is able to predict the best AP position in a complex 3D environment and minimize the failure probability in the robot path planning task. Finally, we discuss immersive experience with 3D radio maps and enabled applications.

\subsection{Performance of URF-GS}

\subsubsection{Experiment Design}

\textbf{Dataset.} We use an open-source dataset from \cite{liang2024gs} and \cite{gentile2024context}. It includes channel measurements operating at  60 GHz  in a high-fidelity digital replica of a real indoor scene from the National Institute of Standards and Technology (NIST). This scenario contains a $14 \times 15 \ \mathrm{m}^{2}$ loft with tables, sofas and televisions, as shown in Fig.~\ref{SynSS}. All objects and elements are physically assigned electromagnetic properties, e.g., conductivity and permeability, enabling reliable electromagnetic propagation simulations.

To evaluate the location generalization capability of the constructed 3D radio maps, we collect channel measurements at multiple Rx locations under various Tx position settings using Sionna RT~\cite{hoydis2023sionna}. Both the Tx and Rx employ planar antennas with isotropic patterns and vertical polarization, which produces high-resolution channel impulse responses. They are then processed to estimate key multi-path components (MPCs), including path loss, delay, angle of departure, and angle of arrival~\cite{papazian2016calibration}.
%To generate the spatial spectrum, i.e., 3D radio map, we utilize a synthetic array mode, which emulates high spatial resolution by aggregating measurements from a small number of stationary or moving antennas. This approach  
Based on these MPCs, we reorganize the spatial spectrum into pinhole-camera images via equirectangular projection as detailed in \cite{wen2024neural}. To validate the generalization ability to new Tx positions within the same scene, we constructed a series of synthetic datasets corresponding to different Tx locations (P1, P2, and P3), as depicted in Fig.~\ref{SynSS}. We adopted varying training samples to evaluate the system performance under different training strategies:
\begin{itemize}
\item \textbf{Full-Training:} We generate a dataset of 3,200 samples {for Tx position P1}, where 80\% (2,560 samples) are used for training and 20\% (640 samples) for testing. Each sample consists of the position coordinates of an arbitrarily selected Rx in space and the corresponding MPCs spectrum, both acquired under a specific Tx location.
\item \textbf{Few-Shot:} For the new Tx position P2, we construct a separate dataset that contains only approximately 10 samples to evaluate the model's few-shot learning capability.
\item \textbf{Zero-Shot:} For the new Tx position P3, we adopt a zero-shot setting where no training sample from this position is provided, and the model is evaluated directly on the test set of this independently simulated dataset.
\end{itemize}
\textbf{Baselines.} We compare URF-GS with three baselines:
\begin{itemize}
    \item\textbf{NeRF\textsuperscript{2}}~\cite{zhao2023nerf2}: NeRF\textsuperscript{2} applies the NeRF technique for spatial spectrum construction.
    It can synthesize the spatial spectrum in scenarios where either the Tx or the Rx is fixed while the other moves to arbitrary positions.
    \item\textbf{RF-3DGS}~\cite{liang2024gs}: RF-3DGS is a fast and efficient method for radio radiance field reconstruction based on 3D-GS. It can obtain the spatial spectrum by encoding multi-modal channel characteristics using spherical harmonic functions.
    \item\textbf{WRF-GS+}~\cite{wen2024neural}: WRF-GS+ presents a high-quality spatial spectrum synthesis method based on 3D-GS. It decouples the radio radiation field into large and small-scale components using a deformable network, while balancing rendering performance and rendering speed.
\end{itemize}

\textbf{Metrics.} We consider three performance metrics:
The peak signal-to-noise ratio (PSNR) measures the fidelity score derived from mean squared error between a reference image and a test image; the
structural similarity index measure (SSIM) \cite{wang2004image} evaluates the perceptual similarity by comparing luminance, contrast, and structure locally; the normalized mean square error (NMSE) \cite{xiao2026scalable} quantifies the relative reconstruction error in terms of signal power, serving as a standard and physically interpretable metric in wireless channel modeling.

\subsubsection{Experiment Result}

\begin{table}[!t] 
  \centering  
  \caption{\textbf{Performance comparison of the URF-GS, RF-3DGS, NeRF\textsuperscript{2}, and WRF-GS+ methods.}}  
  \label{tab:performance} 
  \begin{tabular*}{0.96\textwidth}{@{\extracolsep{\fill}}lcccc@{\extracolsep{\fill}}}  
    \toprule  
    \textbf{Training Data} & \textbf{Method} & \textbf{PSNR $\uparrow$} & \textbf{SSIM $\uparrow$} & \textbf{NMSE $\downarrow$}  \\
    \midrule  
    \multirow{4}{*}{\textbf{Full-Training}} 
    & URF-GS (Ours) & \textbf{17.3818} & \textbf{0.7012} & \textbf{0.0615} \\
    & RF-3DGS~\cite{zhang2024rf} & 16.9949 & 0.6823 & 0.0668 \\
    & NeRF$^2$~\cite{zhao2023nerf2} & 17.0581 & 0.5623 & 0.0658 \\
    & WRF-GS+~\cite{wen2024neural}   & 17.1085 & 0.6917 & 0.0649 \\
    \midrule
    \multirow{4}{*}{\textbf{Few-Shot}} 
    & URF-GS (Ours)  & \textbf{12.5612} & \textbf{0.5035} & \textbf{0.1863} \\
    & RF-3DGS & 8.4301 & 0.3358 & 0.4824 \\
    & NeRF$^2$ & 7.3301 & 0.3876 & 0.6216 \\
    & WRF-GS+   & 10.9804 & 0.4859 & 0.2680 \\
    \midrule
    \multirow{4}{*}{\textbf{Zero-Shot}} 
    & URF-GS (Ours) & \textbf{9.3316} & \textbf{0.3156} & \textbf{0.3927} \\
    & RF-3DGS & 7.0124 & 0.2794 & 0.6765 \\
    & NeRF$^2$ & 6.5123 & 0.2846 & 0.7596 \\
    & WRF-GS+   & 7.3512 & 0.2946 & 0.6180 \\
    \bottomrule 
  \end{tabular*}
\end{table}

\begin{figure*}[!t]
\centering
    \subfloat[PSNR.]{\includegraphics[width=0.33\linewidth]{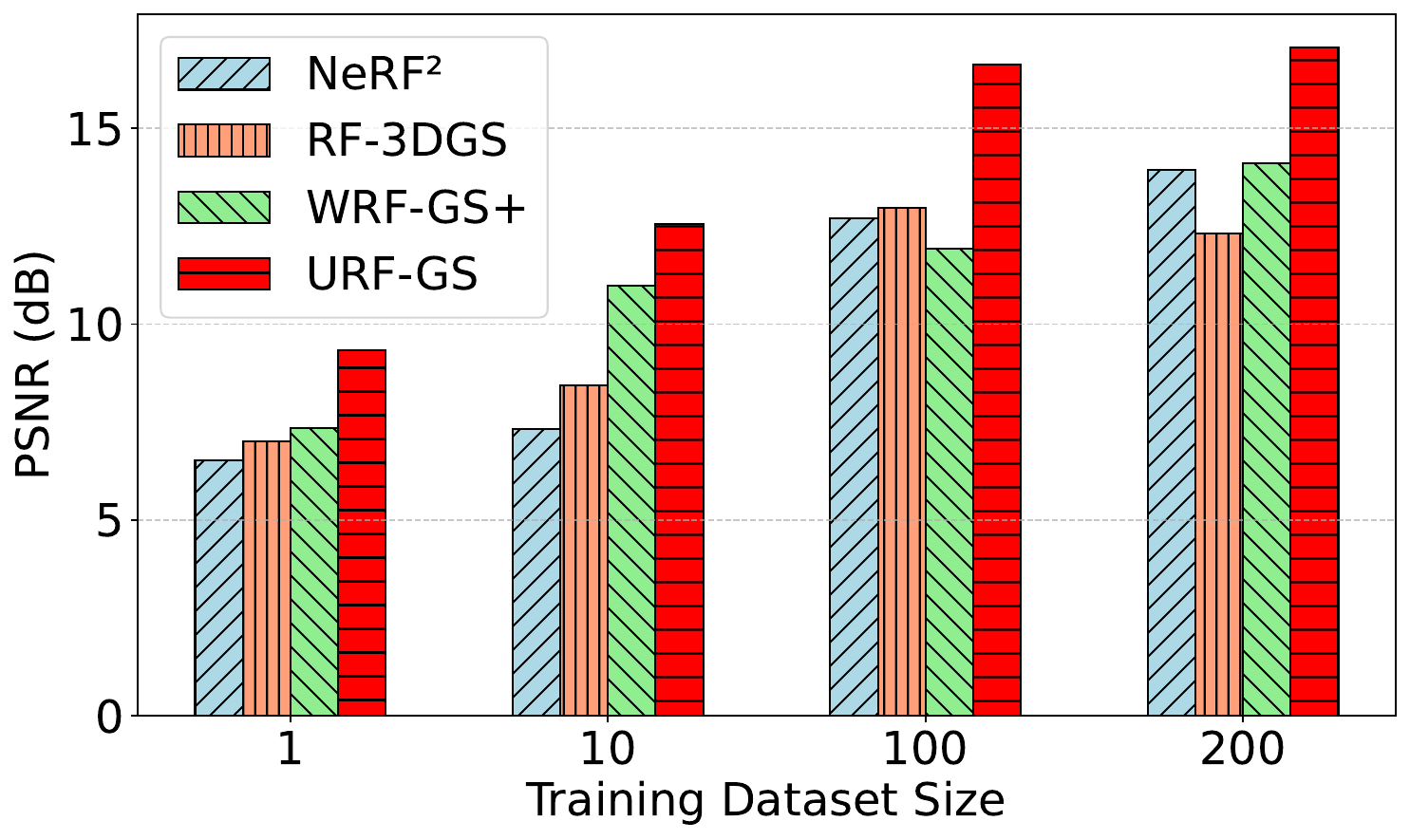}}
    \subfloat[SSIM.]{\includegraphics[width=0.33\linewidth]{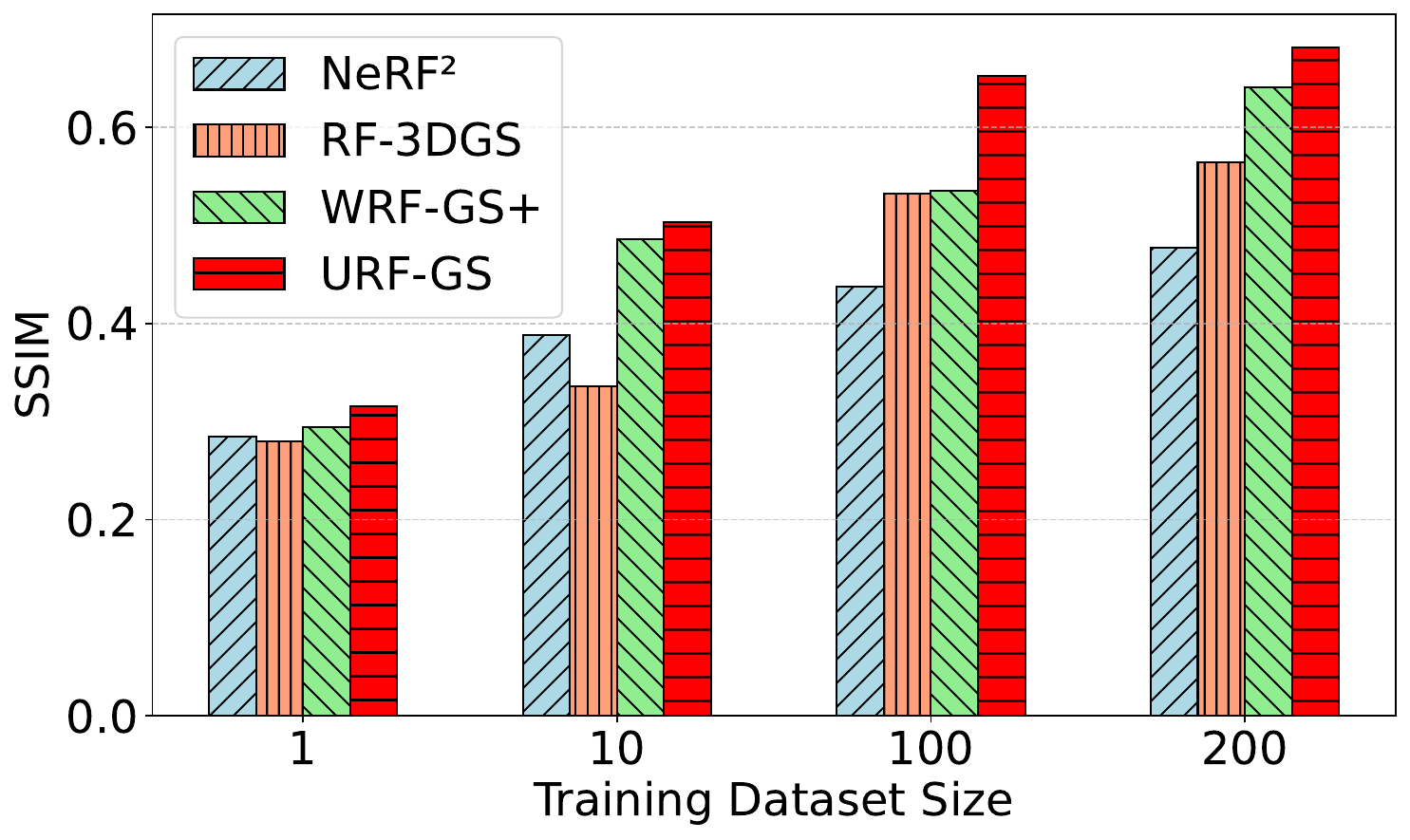}}
    \subfloat[NMSE.]{\includegraphics[width=0.33\linewidth]{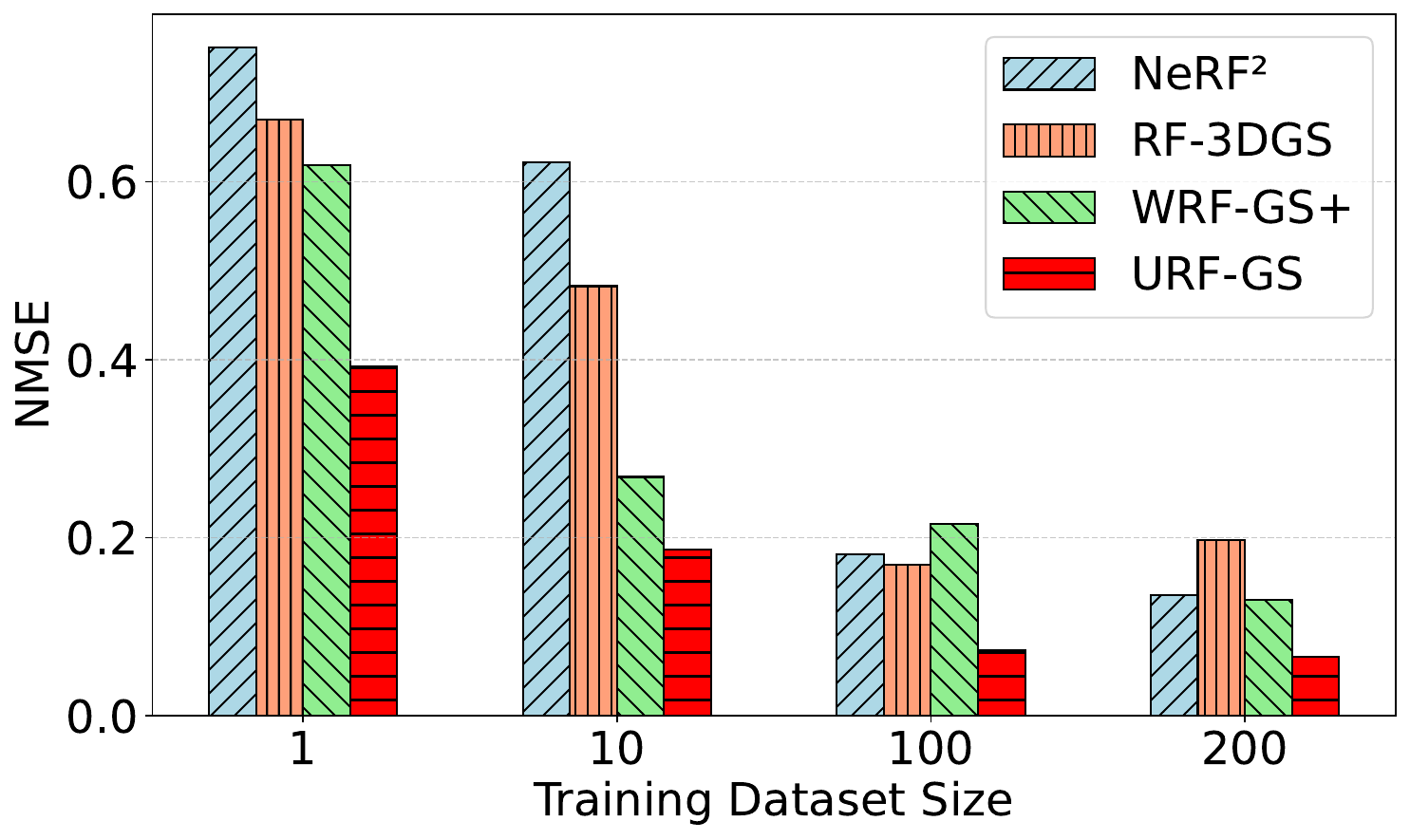}}
    \caption{\textbf{Performance comparison of the URF-GS, RF-3DGS, NeRF\textsuperscript{2}, and WRF-GS+ methods in terms of PSNR, SSIM, and NMSE metrics over different training dataset sizes.}}
    \label{METRIC}
\end{figure*}

Table~\ref{tab:performance} compares URF-GS with RF-3DGS, NeRF\textsuperscript{2}, and WRF-GS+ in terms of PSNR, SSIM, and NMSE.
We see that when we train the model with sufficient data, URF-GS achieves the best PSNR (17.3818) and SSIM (0.7012), with an improvement of up to $\mathbf{24.7}\%$ in SSIM compared with NeRF$^2$. It also attains the lowest NMSE (0.0615), indicating the smallest relative reconstruction error among all methods. While WRF-GS+ achieves a competitive NMSE of 0.0649, it is outperformed by URF-GS in PSNR and SSIM, suggesting that URF-GS delivers more accurate and structurally coherent reconstructions. NeRF$^2$ trails notably in SSIM (0.5623). Overall, URF-GS demonstrates consistent superiority across all evaluation metrics, underscoring the advantage of the GS-based design tailored for wireless radiance field reconstruction. Furthermore, we quantitatively evaluate the generalization performance of various methods in the few-shot (10 samples) and zero-shot training scenarios after adjusting the position of TX in the same scene. The results indicate the superiority of the proposed URF-GS scheme among all evaluation metrics.

Fig.~\ref{METRIC} further evaluates the generalization capability of URF-GS with respect to Rx locations in the few-shot training setup. We compare URF-GS with $\text{NeRF}^2$, RF-3DGS, and WRF-GS+ with increasing training set sizes in terms of PSNR, SSIM, and NMSE. For all dataset sizes, URF-GS achieves the best performance. It consistently yields the highest PSNR and SSIM values, while attaining the lowest NMSE, reflecting superior reconstruction fidelity and physical accuracy. Notably, the gains are most pronounced in low-data settings (1–10 samples), which achieves up to $\mathbf{10\times}$ sample efficiency compared with the NeRF\textsuperscript{2} method, demonstrating a high sample efficiency for the proposed unified pipeline. The advantage persists and even widens at larger scales (100–200 samples), suggesting a better capacity to leverage additional data without overfitting. Overall, the results validate that URF-GS delivers high-quality 3D radio map construction and a strong location generalization capability.

\vspace{0.5cm}
\noindent \textbf{Computational Efficiency.} We further evaluate the inference efficiency of URF-GS, as a representative of 3DGS-based methods, against two other paradigms: NeRF$^2$ (implicit neural field) and Sionna RT (conventional ray-tracing). All tests are conducted on a single NVIDIA RTX 3090. Sionna RT uses the exact settings employed for our ground-truth generation. Table~\ref{tab:efficiency} reports the average latency and peak GPU memory for synthesizing a single $200\times300$ AoA spectrum from a trained model.

\begin{table}[h]
\centering
\caption{Inference efficiency on a single RTX 3090.}
\label{tab:efficiency}
\begin{tabular}{lcc}
\toprule
\textbf{Method} & \textbf{Latency (ms / spectrum)} & \textbf{Peak GPU Mem. (MB)} \\
\midrule
Sionna RT~\cite{hoydis2023sionna} & 368 & 324 \\
NeRF$^2$~\cite{zhao2023nerf2} & 789.9 & 2,712 \\
URF-GS (Ours) & \textbf{11.1} & \textbf{238} \\
\bottomrule
\end{tabular}
\end{table}

URF-GS is \textbf{33$\times$} faster than Sionna RT and \textbf{71$\times$} faster than NeRF$^2$, with the smallest peak GPU memory footprint. This advantage comes from two design choices. First, 3D-GS relies on parallel rasterization, avoiding the dense per-ray sampling of NeRF$^2$. Second, the G-buffer pre-stores geometry and visibility, so the radio field computation can reuse these buffers without repeating ray-scene queries. Sionna RT, even with NVIDIA's hardware acceleration, still runs a full path-tracing pass for each receiver location. Overall, these results show that URF-GS combines high reconstruction quality with real-time inference speed and low memory cost.

\vspace{0.5cm}
\noindent \textbf{Ablation Study.} To quantify the contribution of each geometric prior in URF-GS, we conduct an ablation study on the Full-Training setting. We compare the full model against the following variants: (i) removing the normal smoothness term, (ii) removing normal supervision entirely, (iii) removing depth supervision entirely, and (iv) removing both depth and normal supervision. The results are reported in Table~\ref{tab:ablation}.

\begin{table}[!t]
\centering
\caption{Ablation study on geometric priors (Full-Training).}
\label{tab:ablation}
\begin{tabular}{lccc}
\toprule
\textbf{Configuration} & \textbf{PSNR $\uparrow$} & \textbf{SSIM $\uparrow$} & \textbf{NMSE $\downarrow$} \\
\midrule
w/o Depth \& Normal & 14.25 & 0.541 & 0.1235 \\
w/o Depth Supervision & 15.40 & 0.603 & 0.0947 \\
w/o Normal Supervision & 16.52 & 0.659 & 0.0726 \\
w/o Normal Smoothness & 17.20 & 0.693 & 0.0638 \\
Full URF-GS & \textbf{17.38} & \textbf{0.7012} & \textbf{0.0615} \\
\bottomrule
\end{tabular}
\end{table}

\noindent Removing depth supervision causes the largest performance drop, confirming that accurate surface positioning is essential for both geometry and subsequent radio field modeling. Normal supervision provides a moderate improvement, particularly in SSIM, reflecting its role in stabilizing surface orientation. The smoothness term has a minor auxiliary effect. Overall, the full model achieves the best results across all metrics, and the progressive degradation as priors are removed indicates that each component contributes positively to the final reconstruction quality.

\subsection{Case Study 1: Wi-Fi Access Point Deployment}
The deployment of Wi‑Fi APs involves determining the optimal placement of APs to achieve the desired coverage and capacity, while accounting for signal propagation, interference, and user distribution. Traditional planning methods rely heavily on site surveys and iterative measurements \cite{shen2017dmad}, which require dense RSSI sampling and manual tuning that are labor-intensive and time-consuming. Moreover, these approaches are sensitive to environment changes and frequently yield suboptimal performance. In contrast, URF‑GS addresses these challenges by learning a unified radiation field that jointly models both the optical scene and radio propagation. This unified representation allows site evaluations where collected signal data closely matches simulation predictions, thereby achieving rapid AP placement without extensive surveys.

\subsubsection{Experiment Design} We evaluate URF-GS in a complex and realistic indoor environment using the publicly available ``Bistro'' scene dataset~\cite{Lumberyard2017Amazon}. The Bistro scene features a fully furnished café interior, with detailed architectural elements such as walls, tables and chairs. This scene also introduces numerous occlusions, multi-path reflections, and shadow zones, making it a highly challenging scenario for accurate channel modeling.

\begin{figure}[!t]
\centering
    \subfloat[Bistro.]{\includegraphics[width=0.43\linewidth]{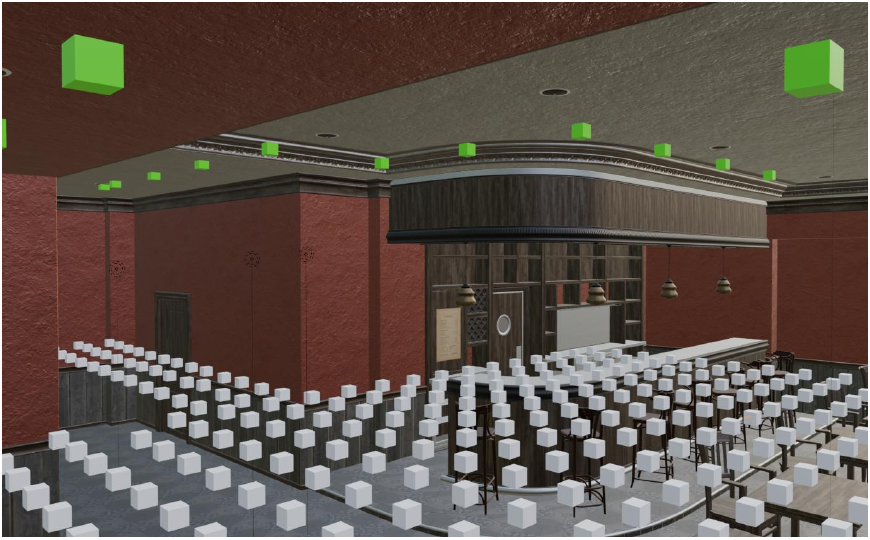}}
    \subfloat[Modified Wi3room.]{\includegraphics[width=0.5\linewidth]{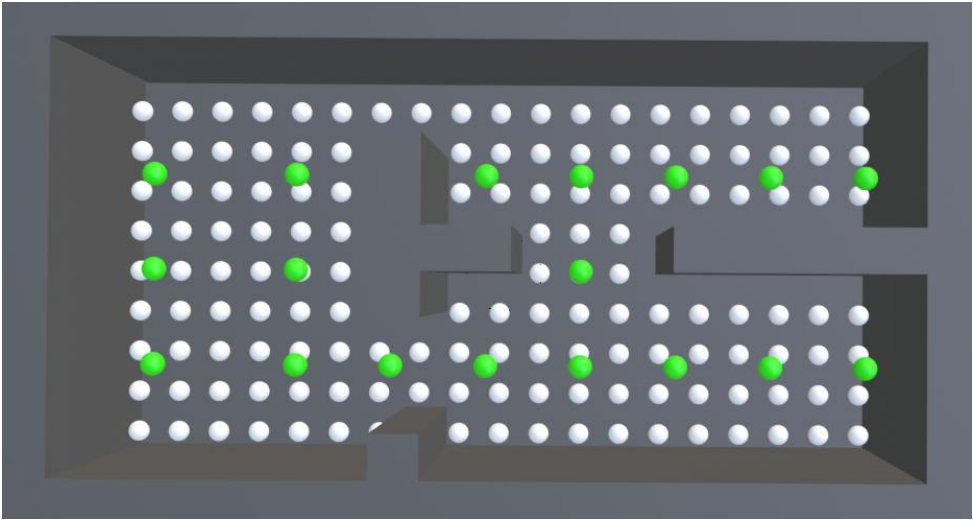}}
    \caption{\textbf{Datasets.} (a) Bistro. The green and white cubes represent the positions of transmitters and receivers, respectively. (b) Modified Wi3room. The green and white spheres represent the positions of the transmitters and receivers, respectively.}
    \label{fig:dataset}
\end{figure}

For training, we randomly select $5$ Tx positions. The testing set consists of $25$ previously unseen transmitters mounted along the ceiling to reflect practical AP deployment. For each Tx, $386$ Rx positions, corresponding to a dense user population, are uniformly sampled at a height of $1.2$ m, which is  the typical height of a user device. Figure~\ref{fig:dataset} visualizes the layout of the scene and the distribution of Tx and Rx. This setup provides a rigorous benchmark for evaluating the generalization capability of URF-GS to novel transmitter placements and its ability to accurately model radio propagation in dense and complex environments.

\begin{figure}[!t]
    \centering
    \includegraphics[width=0.6\linewidth]{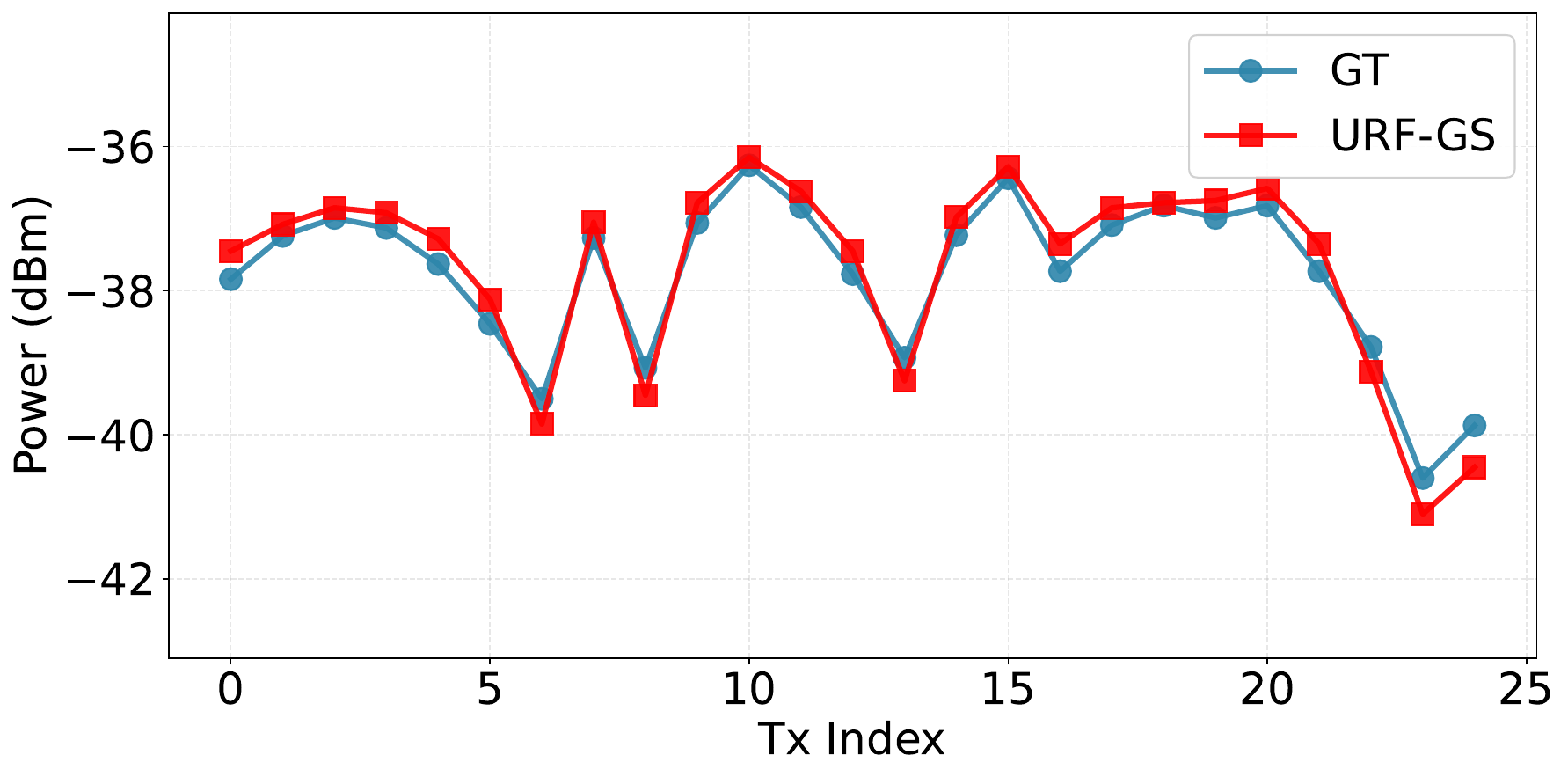}
    \caption{\textbf{The average received power of all receivers at different Tx positions.}}
    \label{fig:AveragePower}
\end{figure}
\begin{figure}[!t]
    \centering
    \begin{subfigure}{0.47\linewidth}
        \centering
        \includegraphics[width=\linewidth]{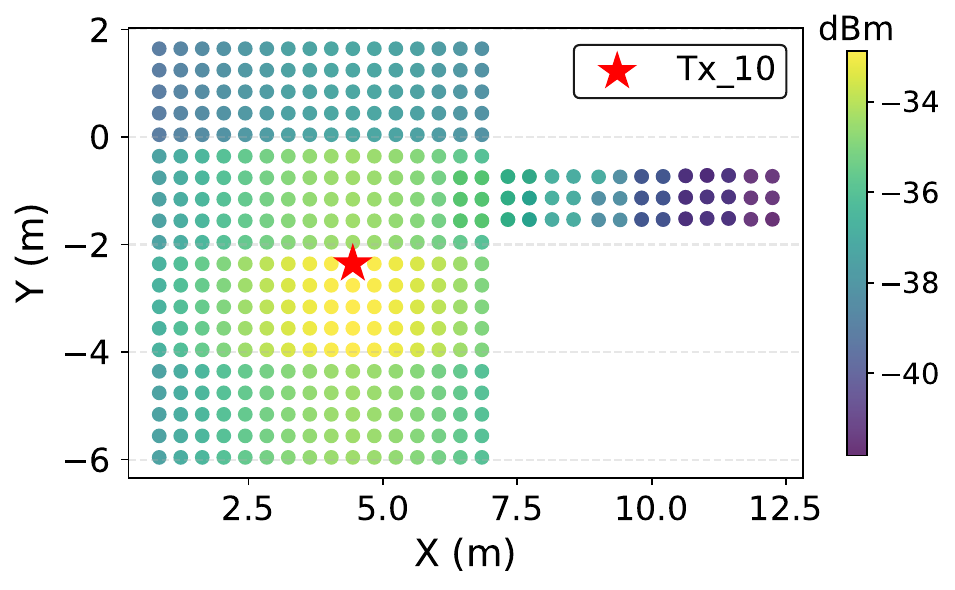}
        \caption{Best GT Tx position.}
    \end{subfigure}
    \begin{subfigure}{0.47\linewidth}
        \centering
        \includegraphics[width=\linewidth]{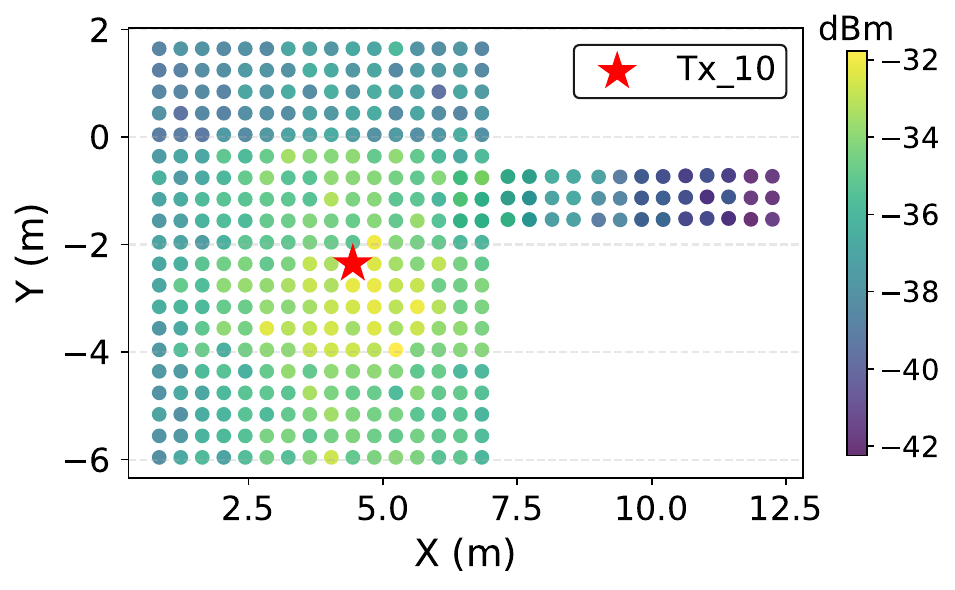}
        \caption{Predict Tx position.}
    \end{subfigure}
    \begin{subfigure}{0.47\linewidth}
        \centering
        \includegraphics[width=\linewidth]{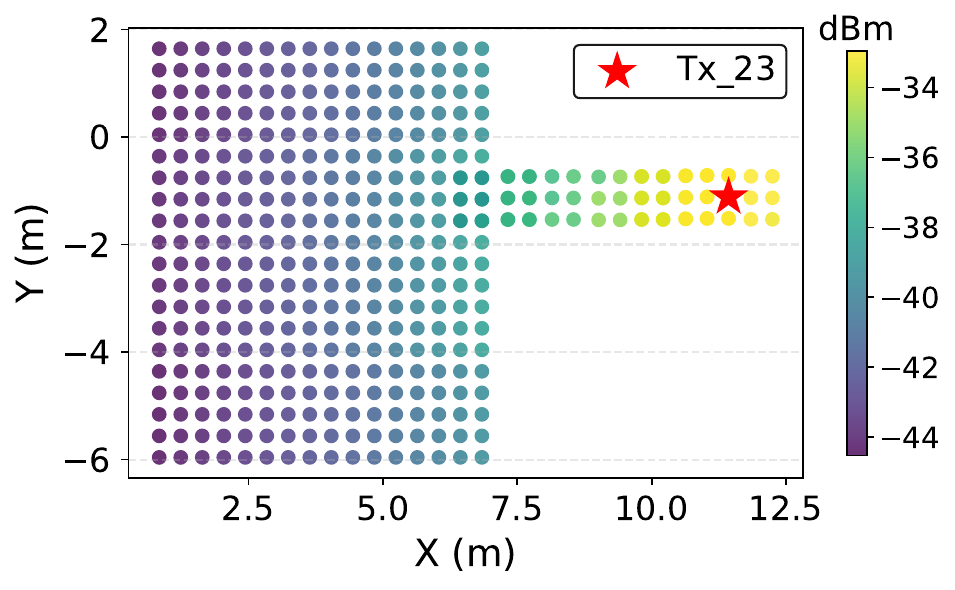}
        \caption{Worst GT Tx position.}
    \end{subfigure}
    \begin{subfigure}{0.47\linewidth}
        \centering
        \includegraphics[width=\linewidth]{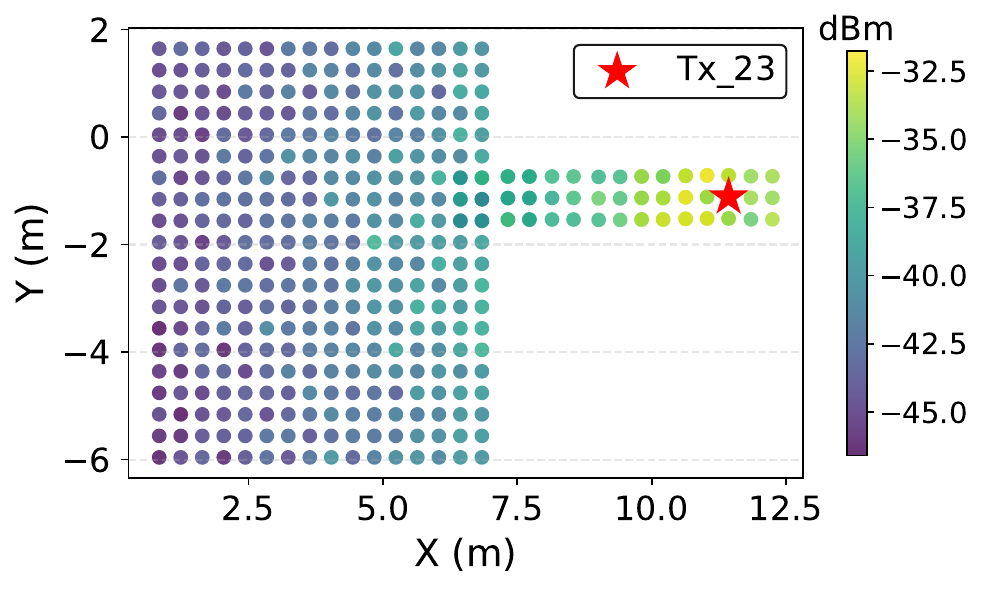}
        \caption{Predict Tx position.}
    \end{subfigure}
    \caption{A visualization of the GT and predicted Tx positions on the indices $10$ and $23$ in the Bistro scene, where the coordinate projection transformation is used in the projection model.}
    \label{Proj}
\end{figure}
\subsubsection{Experiment Result} 
Fig.~\ref{fig:AveragePower} compares the average received power across all receivers for different transmitter positions predicted by the proposed URF-GS with GT. The two curves closely match each other over all $25$ Tx indices in the testing dataset, capturing both global trends and local variations caused by blockage and MPCs. URF‑GS slightly overestimates power in some cases, e.g., around Tx 1–3 and 14–16, and underestimates near the weakest links (Tx 22–24), but the deviations are small relative to the overall dynamic range. It is observed that the Tx positions with indices $10$ and $23$ lead to the best and worst average received power, respectively. As shown in Fig.~\ref{Proj}, we plot the Tx positions with indices $10$ and $23$, as well as the corresponding received power. It is seen that for both Tx positions, the predicted received powers match well with the GT in both situations. This demonstrates that URF‑GS can reliably rank candidate AP locations and approximate their coverage quality without exhaustive measurements, providing a practical basis for data‑efficient AP planning. 

\subsection{Case Study 2: Robot Path Planning}
Robot path planning focuses on computing collision-free and task-efficient trajectories for mobile robots given environmental constraints, such as safety, time, and energy. Classical and learning-based planners primarily exploit optical geometry from LiDAR/cameras to build maps and avoid obstacles. However, they often ignore the radio propagation and connectivity that affect communication, teleoperation reliability, and data offloading. URF-GS addresses this problem by providing a unified 3D radio map to augment geometric maps for robot path planning. This enables efficient robot navigation that jointly optimizes path planning and adapts to the new Tx positions with strong location generalization.

\subsubsection{Dataset} 
We generate a synthetic dataset based on the Wi3room scene~\cite{orekondy2023winert} by removing the wall between the left room and the upper right room. We place only one single channel that connects the three rooms. Such modifications are used to construct a challenging robot path planning task.
Specifically, there are $110$ positions for Rx uniformly selected at a height of $0.2$ m in the scenario, as illustrated in Fig.~\ref{fig:dataset}. This setting is used to mimic the height of the receiving antennas in a typical navigation robot. Meanwhile, there are $18$ Tx positions uniformly arranged on the ceiling to ensure that the radio signals could cover each room. The objective is to minimize the failure probability of the planned path using the geometry information and the radio map.

\subsubsection{Experiment Design}
We consider the Wi3room environment with size $19 \times 8$ m  which is further divided into $1\times 1$ m grids for the robot path planning task. In each grid, the robot can stop or move in four directions, where the Start, Goal, and Tx positions are randomly placed. We define the mission failure probability $P$ as the event in which the average signal power of a grid reached by the robot falls below a predefined power threshold. Both the power threshold and the probability threshold can be adjusted. To discourage excessively long detours for a better signal, we further set the maximum step count constraint, reflecting real energy constraints. In simulation, we first train a URF‑GS model on Tx–Rx data from this Wi3room environment. The reconstructed URF is used for path planning. Note that the trained model can generalize to new Tx placements with zero-shot samples and predict signal power at any Rx location.

\begin{figure}[!t]
    \centering
    \includegraphics[width=0.7\linewidth]{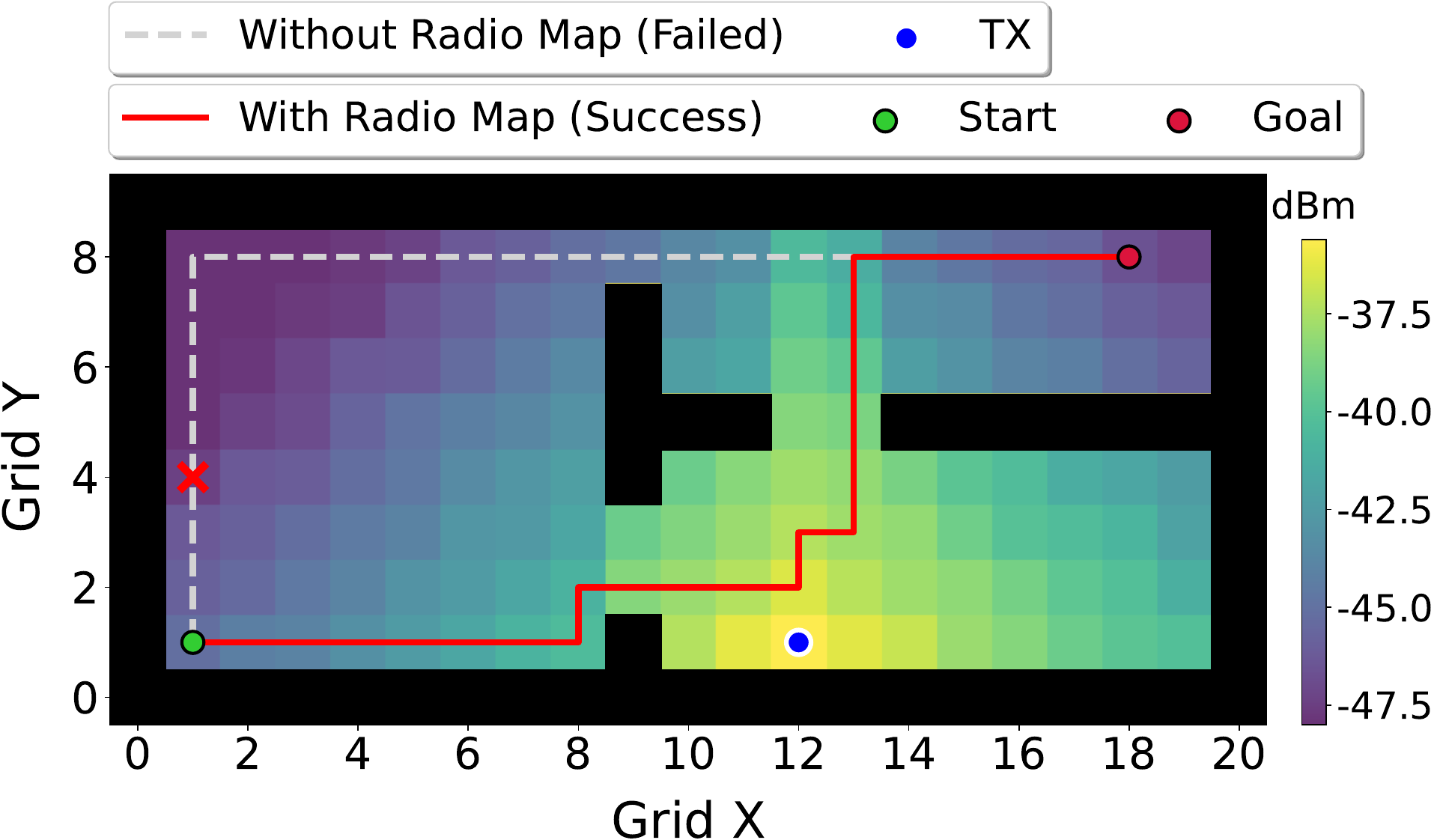}
    \caption{\textbf{Comparison of path planning results with and without radio map assistance.}}
    \label{fig:PathPlan2}
\end{figure}
\begin{figure}[!t]
    \centering
    \includegraphics[width=0.7\linewidth]{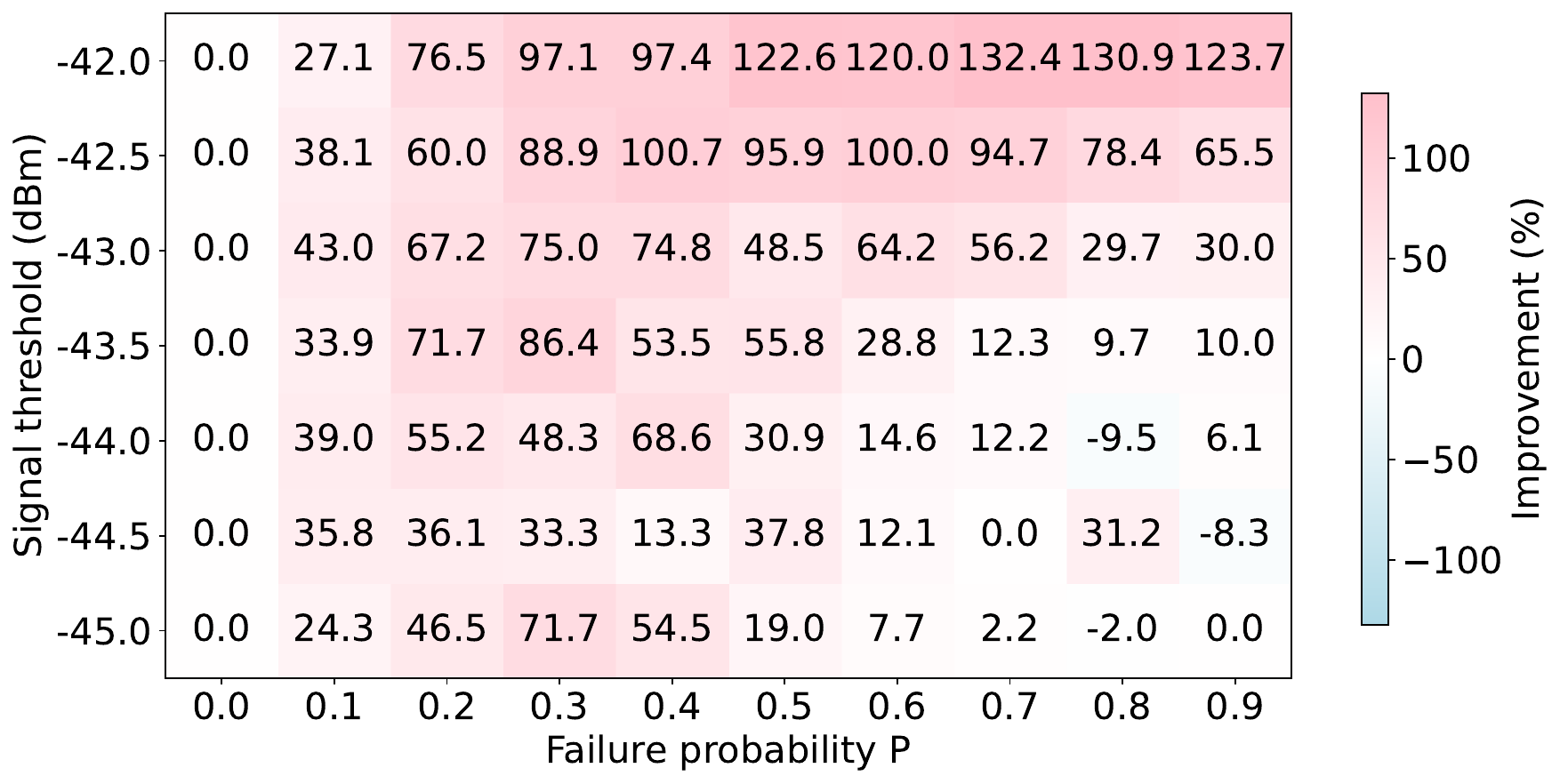}
    \caption{\textbf{The improvement rates of the 3D radio map-enabled path planning method under different signal power thresholds and failure probabilities.}}
    \label{fig:improvement}
\end{figure}
\subsubsection{Experiment Result}
Fig.~\ref{fig:PathPlan2} depicts the planned robot paths with and without the URF‑GS radio map. The baseline (gray dashed) follows the shortest visual route from Start to Goal, but it traverses low‑power regions (purple), leading to failure due to communication‑reliability constraint. In contrast, the radio‑aware planner (red) detours through higher‑power areas (green/yellow) while circumventing obstacles, maintaining connectivity and successfully reaching the goal. This demonstrates that incorporating the learned radio map enables connectivity‑aware navigation that can optimize path selection based on signal quality even when path lengths are similar, leading to improved task success.

Fig. \ref{fig:improvement} shows the improvement rates of the proposed method under different signal power thresholds and failure probabilities. We see that the proposed path planning method shows zero improvement when the failure probability, $P = 0$. However, its performance gains increase significantly as failure probability rises: a $132.4\%$ improvement is observed when  $P=0.7$ and the received power threshold equals to $-42.0$ dBm.
The proposed method is more effective under stricter signal constraints, consistently maintaining high improvement rates (above $97\%$) when the threshold equals to $-42.0$ dBm.
In contrast, performance gains diminish at lower power thresholds, e.g., $-45.0$ dBm, and extremely high failure rates, occasionally resulting in slight performance degradation. This result demonstrates that the proposed method can effectively minimize the failure probability of the planned path by using the geometry information and the radio map.

\subsection{Immersive Experience of 3D Radio Map}
We provide an immersive demonstration of the 3D radio map using a VR device, employing Meta Quest 3 as the head-mounted display. The trained Gaussian model is uploaded to a processing system, which transmits the URF-GS to Meta Quest 3 via a web interface~\cite{GaussianSplats3D}, enabling real-time rendering and interactive exploration. Notably, URF-GS utilizes spherical harmonic functions to encode optical color information, while attributes such as albedo, roughness, and metallicity are used to represent radio-related material properties. By simply modifying the device shader, users can seamlessly switch between rendering the optical radiation field and the wireless radiation field from the same viewpoint.

\begin{figure}[!t]
    \centering
    \includegraphics[width=0.5\linewidth]{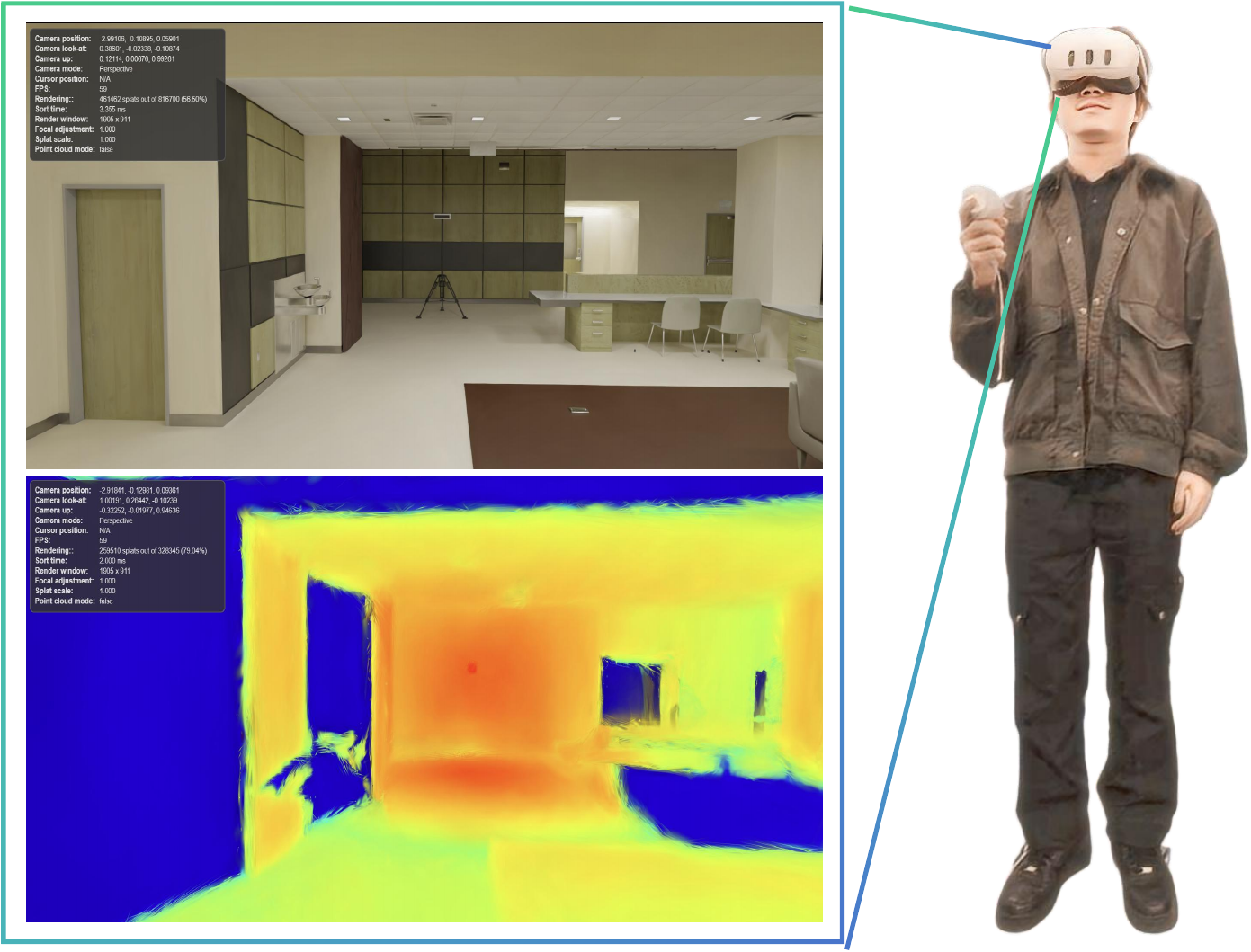}
    \caption{\textbf{Immersive Experience of URF with VR Devices. }}
    \label{fig:VR}
\end{figure}

As shown in Fig.~\ref{fig:VR}, both the 3D optical scene and the radio map can be visualized within Meta Quest 3 based on the reconstructed URF. Users can move freely within the scene and adjust their perspective to experience the URF in an immersive way, with intuitive toggling between optical and wireless views. Benefiting from the unified 3D Gaussian-based representation and the efficiency of 3D-GS for real-time rendering, the system maintains a stable frame rate of approximately $\mathbf{60}$ frames per second (FPS) on Meta Quest 3.

Beyond visualization, the immersive 3D radio map holds transformative potential for next-generation network planning and digital twin applications. By enabling engineers to ``see" invisible wireless signal propagation overlaid on physical environments in real time, this technology could revolutionize site surveys for 5G and 6G deployment, allowing for precise identification of coverage dead zones and interference sources without complex simulations. Furthermore, in the realm of robotics and autonomous systems, integrating such granular radio maps could facilitate ``connectivity-aware" navigation, where drones or autonomous vehicles dynamically adjust their paths to maintain optimal link quality in cluttered urban environments. Ultimately, this unified representation of optical and wireless data bridges the gap between the physical and digital worlds, paving the way for more resilient and intelligent smart city infrastructures.

\section{Discussion}\label{sec3}

This paper introduced URF-GS, a unified and generalizable framework for 3D radio map construction by leveraging 3D-GS and physics-informed inverse rendering. URF-GS achieves an accurate and physically consistent reconstruction of scene geometry, material properties, and radio signal propagation by jointly modeling optical and radio-frequency radiation fields through a unified Gaussian representation. Extensive experiments demonstrated that URF-GS consistently outperforms state-of-the-art methods in both accuracy and generalization ability, and its versatility in tasks like Wi‑Fi AP deployment and robot path planning underscores its practical value for next‑generation wireless networks.

The design of URF-GS is motivated by fundamental limitations of existing paradigms. Conventional ray-tracing approaches demand precise geometric models and explicit material assignments that are rarely obtainable in real-world deployments. Even when approximate Computer-Aided Design (CAD) models are available, the manual effort and calibration required make them impractical for dynamic or large-scale scenarios. In contrast, URF-GS extracts scene geometry directly from visual data, bypassing the need for hand-crafted meshes while still capturing the essential spatial structures that govern wave propagation. Compared with 2D or 2.5D methods that rely on top-down views or grid-based regression, our explicit 3D representation naturally encodes height variations, occlusions, and arbitrary perspectives, offering a more complete and actionable environmental model. Furthermore, unlike purely implicit neural approaches that often struggle to generalize beyond their training distribution, URF-GS grounds its predictions in explicit, view-consistent geometry. This fusion of visual priors with wireless measurements provides a strong inductive bias, enabling robust generalization to unseen transmitter-receiver configurations and making the framework resilient in the few-shot and zero-shot regimes evaluated in this work.

While URF-GS demonstrates strong performance in static and well-defined environments, several open challenges remain. First, the framework currently assumes static scenes and does not account for moving objects or humans. Extensions to dynamic 3D-GS representations would significantly broaden its applicability to more realistic and complex wireless environments. Second, URF-GS learns material-related parameters that are implicitly tied to a specific frequency band and lacks explicit cross-frequency generalization. Extending the framework to handle multiple frequencies would require modeling how materials respond across the spectrum. This also relates to challenging surface types such as transparent or specular objects, where visual reconstruction is difficult and material behavior dominates the radio interaction. Future work could explore frequency-aware material models, possibly guided by semantic priors from vision models. Third, transferring the URF-GS framework across different scenes remains challenging. Promising directions include pretraining on large, multi-scene datasets and adopting continual learning to adapt to environmental changes. Finally, the current model is trained on synthetic ray-tracing data. A natural next step is sim-to-real transfer by fine-tuning the learned representation on measured channel data, bridging the gap between simulation and practical deployment.

\section{Methods}\label{sec4}
In the following, we delve into the technical details of URF-GS. It is designed to construct an accurate 3D radio map that captures both the optical scene and radio signal characteristics by leveraging 3D-GS and inverse rendering. It introduces a unified representation of the radio–optical radiation field, where the optical component reconstructs the visual scene and the radio component synthesizes the spatial spectrum, characterizing the spatial distribution of signal power as it propagates through the environment. To further enhance accuracy and generalizability, URF-GS employs physics-informed inverse rendering to jointly optimize material properties and radiation patterns from multimodal observations, enabling accurate 3D radio map construction under arbitrary Tx–Rx configurations. 

\begin{figure*}[!t]
    \centering
    \includegraphics[width=0.98\linewidth]{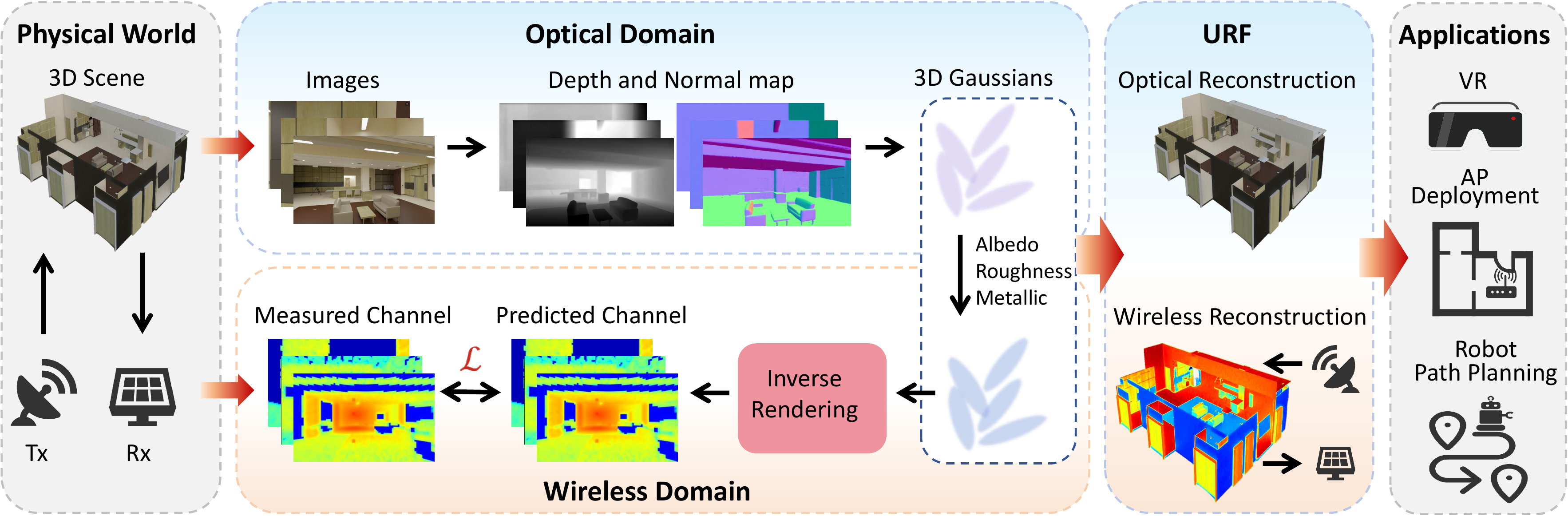}  
    \caption{\textbf{An overview of URF-GS.} From left to right, the modules are the visual and wireless sensing inputs, the proposed two-step training pipeline, the constructed URF, and downstream applications. First, the optical pipeline captures scene images with commodity devices, applies off-the-shelf monocular methods to estimate depth and normals, and uses these priors to train 3D Gaussians for accurate geometry. Next, the radio pipeline consumes the recovered scene geometry and channel measurements from antenna arrays, performing physics-informed rendering to refine the material-related attributes of the 3D Gaussians. Finally, with a new Tx as input, it can rapidly construct an accurate URF, enabling diverse applications, such as AR/VR, Wi-Fi AP deployment, and robot path planning.}
    \label{fig:Pipeline}
\end{figure*}

The overview of the URF-GS is illustrated in Fig.~\ref{fig:Pipeline}. It inputs visual images and wireless measurements and processes them through a two-step representation pipeline using shared Gaussian primitives for both optical and wireless domains. In the optical domain, a set of scene images are collected for training. Then, it uses the off-the-shelf monocular models to estimate image depth and surface normals. These predictions serve as pseudo ground-truth to supervise the corresponding geometric attributes of the scene’s Gaussians. Empirically, integrating such geometric priors allows the 3D Gaussians to recover depth more accurately and produce smoother, more coherent normals, thereby enhancing multi-view consistency and overall scene fidelity.

In the wireless setup, an antenna array acts as the Rx, capturing signals from the Tx at designated positions. We use equirectangular projection to convert the collected angular power or spatial spectrum into pinhole-camera images. During training, the pipeline ingests both the spatial spectrum and learned geometry. The spectrum supervises the material attributes of the Gaussians, while the geometry remains frozen. Additionally, a physics-informed inverse rendering loop decomposes scattering phenomena to simulate wave propagation. During inference, the trained URF model requires only a location input to reconstruct both the optical image and the spatial spectrum. 

To summarize, URF-GS makes two key innovations:
\begin{enumerate}
\item Unlike vanilla 3D-GS, URF-GS leverages monocular models to extract scene normals and depth priors from the input images, thereby imposing stronger geometric constraints on the 3D Gaussians.
\item Instead of overfitting to scene data and only interpolating sparse samples, URF-GS decomposes the scene’s wireless radiation field using geometric information and spectral measurements. This decomposition allows URF-GS to explicitly learn material-related properties, enabling accurate simulation of signal propagation between arbitrary Tx and Rx positions.
\end{enumerate}
To detail how these innovations are implemented, the next two sections dissect the system along its two complementary pipelines: Section~\ref{Sec_OD} presents the optical representation pipeline, including priors, supervision, and rendering; Section~\ref{Sec_WD} introduces the wireless representation pipeline, covering spatial spectrum extraction, material-aware parameterization, and physics-informed inverse rendering.

\subsection{Optical Domain}\label{Sec_OD}
Optical radiation field reconstruction is the first and foundational step in URF-GS, where 3D-GS is used to encode the geometry of the environment. Accurate geometry is critical for 3D radio map construction, as it strongly influences electromagnetic wave propagation. However, vanilla 3D-GS often exhibits multi-view geometric inconsistencies, lacks surface normal information, and produces visual artifacts. To overcome these issues, we adopt a prior-regularized reconstruction technique~\cite{Yu2022MonoSDF, turkulainen2025dn} to enable high-fidelity optical geometric recovery. Specifically, we first introduce the preliminary knowledge of 3D-GS in Section~\ref{SubSec_3GS}. Building on this foundation, Section~\ref{SubSec_LDC} introduces monocular depth priors and an edge-aware loss to regularize Gaussian positions. Thereafter, Section~\ref{SubSec_LNC} details the extraction of surface normal directions from Gaussians. Finally, Section~\ref{SubSec_GI} presents different strategies for initializing 3D-GS using depth and normal information.

\subsubsection{3D Gaussian Splatting}\label{SubSec_3GS}
3D-GS is an emerging technique for 3D scene reconstruction, which utilizes a set of 3D Gaussian primitives $G(\mathbf{x})$ to model the geometry layout of a scene, offering a compact and explicit representation \cite{kerbl3Dgaussians}. 
3D-GS represents the 3D scene using anisotropic Gaussian primitives, where each Gaussian is characterized by its mean $\mathbf{\mu} \in \mathbb{R}^{3}$ and covariance matrix $\mathbf{\Sigma} \in \mathbb{R}^{3 \times 3 }$ with the probability distribution function defined as
\begin{equation}
    G ( \mathbf{x} )= {e}^{-\frac{1}{2}(\mathbf{x}-\mathbf{\mu} )^{T} \mathbf{\Sigma}^{-1} (\mathbf{x}-\mathbf{\mu}) }.
\end{equation}
%where $\mathbf{x}$ is the spatial coordinates of a Gaussian point in 3D space. 
The covariance matrix $\mathbf{\Sigma}$ can be expressed using the scaling matrix $\mathbf{S}$ and the rotation matrix $\mathbf{R}$ as follows:
\begin{equation}
    \mathbf{\Sigma} =\mathbf{R}\mathbf{S}\mathbf{S}^{T}\mathbf{R}^{T}.
\end{equation}

The rendering process begins by projecting the 3D Gaussians $G(\mathbf{x})$ onto the image plane, resulting in 2D Gaussians, denoted by $G'(\mathbf{x'})$  \cite{zwicker2002ewa}. 
%These 2D Gaussians are then sorted based on their depth information to ensure proper handling of occlusions. 
%Each tile on the image plane is processed independently and in parallel for the 2D Gaussians within its coverage.
Then, $\alpha$-blending operation is performed where the opacity ${o}_{i}$ and the color attribute $c_i$ of each Gaussian are utilized to compute each of the pixel value of the rendered image, i.e.,
\begin{equation}
    C = \displaystyle\sum_{i=1}^{N}{T_i}{\alpha}_{i}{c}_{i},\quad \textrm{with} \quad T_i = \prod_{j=1}^{i-1}(1-{\alpha}_{j}),
\label{eq:redering0}
\end{equation}
where ${\alpha}_{i}={o}_{i}{G}_{i}^{'}(\varDelta \mathbf{p}_{i})$, and $\varDelta \mathbf{p}_{i}=\mathbf{p}_{i}-\mathbf{p}_{s}$ represents the difference between the center of a projected 2D Gaussian $\mathbf{p}_{i}$ and the pixel position $\mathbf{p}_{s}$. 
Moreover, $N$ denotes the number of 3D Gaussians that contribute to a given pixel.
The rendered image is obtained after performing the aforementioned $\alpha$-blending operation for all pixels. The 3D-GS parameters are optimized by calculating the pixel-wise loss between the rendered image and the GT.
%The rendered image is then compared with the ground truth to calculate the pixel-wise loss and used to optimize the model parameters.

\subsubsection{Utilizing Depth Cues}\label{SubSec_LDC}
Next, we introduce the monocular depth priors and an edge-aware loss to regularize the Gaussian positions. 

\textbf{Depth Prediction.}  Depth information measures the distance between an object and the viewpoint, which is essential to determine the position of the objects and the occlusion relationships between objects. To accelerate rendering, 3D-GS employs tile-based parallel rasterization \cite{zwicker2002ewa}, where each tile independently performs depth sorting of the Gaussians. Since this sorting is performed locally rather than per-pixel along each viewing ray, the resulting depth ordering is only an approximate and may be inconsistent across different tiles and viewpoints. Consequently, this introduces errors in the occlusion relationships between Gaussians, potentially leading to rendering artifacts. To address this limitation, we adopt linear interpolation to compute the depth values of $N$ Gaussians as in GS-IR~\cite{liang2024gs}:
\begin{equation}
    \hat{\textbf{D}}={{\sum_{i= 1}^{N}{\hat{w}_i}}}d_i,\quad \mathrm{with} \quad {\hat{w}}_{i}=\frac{T_i{\alpha_i}}{\sum_{i= 1}^{N}{T_i}{\alpha_i}},
    \label{eqn:depth}
\end{equation}
where $d_i$ denotes the distance from the $i$-th 3D Gaussian to the image plane and $T_i$ is defined in \eqref{eq:redering0}. Compared with the $\alpha$-blending in \eqref{eq:redering0}, this approach reduces artifacts when predicting the depth of each Gaussian, leading to improved rendering performance.
%enabling a smooth transition between the deepest and shallowest Gaussians.
%and ensures a smoother transition in rendered depth values between the deepest and shallowest Gaussians.

\textbf{Monocular Depth Regularization.} 3D-GS renders a depth map without GT when the dataset lacks depth data. However, relying solely on the visual appearance of the scene and geometric information across multiple views can lead to inaccurate depth predictions. To mitigate this issue, we leverage pre-trained monocular models for depth estimation \cite{bhat2023zoedepth, yang2024depth} to obtain dense pixel-wise depth priors. Furthermore, we utilize the structure from motion (SfM) method \cite{schonberger2016structure} to address the scale ambiguity problem in the monocular modeling process by providing sparse yet scale-accurate depth information~\cite{Yu2022MonoSDF}.

Specifically, for each monocular depth map $\textbf{D}_{\text{Mono}}$, we align its scale with that of the sparse depth map $\textbf{D}_{\text{SfM}}$, which is generated by projecting SfM points onto the camera view. This is achieved by solving the per-image scale parameter $a$ and the shift parameter $b$ using the closed-form linear regression solution, i.e., 
\begin{equation}
    \hat{a}, \hat{b} = \arg\min_{a,b} \sum_{m,n} \bigl\| \left(a \cdot D_{\text{Mono}}^{m,n} + b\right) - D_{\text{SfM}}^{m,n} \bigr\|_2^2,
\end{equation}
where \( D_{\text{Mono}}^{m,n}\) and \( D_{\text{SfM}}^{m,n}\) denote the depth values at the \((m,n)\)-th pixel in the depth maps, \(\textbf{D}_{\text{Mono}}\) and \(\textbf{D}_{\text{SfM}}\), respectively. This formulation enforces a linear alignment between the corresponding depth values of the two maps. 

We define the aligned depth map as 
\begin{equation}
    \textbf{D}_{\text{aligned}} \triangleq \hat{a} \cdot \textbf{D}_{\text{Mono}} + \hat{b},
    \label{eq:aligned_depth_map}
\end{equation}
 which will be utilized as the ground truth to supervise \(\hat{\textbf{D}}\) in \eqref{eqn:depth}. 
Then, we employ an edge-aware loss to effectively balance regularization across various image regions while accommodating the geometric and textural complexity of the scene, which is expressed as
\begin{equation}
    \mathcal{L}_{\hat{\textbf{D}}} = \frac{1}{|\hat{\textbf{D}}|} g_{\text{rgb}} \log\left(1 + \|\hat{\textbf{D}} - \textbf{D}_{\text{aligned}}\|_1\right),
\end{equation}
where $\vert \hat{\textbf{D}} \vert$ represents the total number of pixels in $\hat{\textbf{D}ni}$, while $ g_{\text{rgb}} \triangleq \exp(-\nabla I)$ with $\nabla I$ denoting the gradient of the currently aligned RGB image.

\subsubsection{Utilizing Normal Cues}\label{SubSec_LNC}
We illustrate the process of extracting surface normal directions from 3D Gaussians and subsequently use these cues for additional regularization of monocular normals.

\textbf{Normal Prediction.} The normal vector, which is perpendicular to the surface of an object, is essential for calculating the relationship between incident and outgoing light. 
In the context of 3D-GS, providing additional normal information helps align the 3D Gaussian primitives with the scene's geometry, resulting in more photorealistic rendered outputs.
%While depth only encodes distance, normals can differentiate between surfaces of identical depth but distinct orientations, which makes normals a key component of 3D reconstruction and rendering. 
Unlike existing methods \cite{shi2025gir, chen2022tensorf}, which use the direction of the shortest axis of the primitive as the normal direction and naively constrain the primitives to be flat disks, {in URF-GS, we incorporate the normal vector $\mathbf{n}_{i}$ as an additional attribute for the $i$-th Gaussian primitive to avoid performance degradation}. The normal map can then be obtained by using $\alpha$-blending, i.e., 
\begin{equation}
    {\hat{\mathbf{N}} = \sum_{i = 1}^{N} \mathbf{n}_i \alpha_i T_i,}
    \label{noraml}
\end{equation}
where $N$, $\alpha_i$, and $T_i$ are given in Eqn. \eqref{eq:redering0}.

\textbf{Monocular Normal Regularization.} Similar to the depth information illustrated in Section \ref{SubSec_LDC}, additional normal information is needed to train the normal attributes $\{\textbf{n}_i\}$ for each of the 3D Gaussian primitives. To this end, the authors in \cite{gao2024relightable} obtain the normal maps by calculating the gradients of the rendered depth maps. However, as shown in Fig. \ref{fig:normal}, due to the noise in the rendered depth maps, this method yields artifacts, especially in complex scenes. To address this, we utilize monocular normal estimates from the pre-trained Omnidata model \cite{eftekhar2021omnidata} to train the normal attributes of the URF-GS model as they exhibit significantly smoother characteristics. The normal properties of the Gaussians primitives in the proposed URF-GS are optimized using the \(\ell_1\)-loss function as:
\begin{equation}
    { \mathcal{L}_{\mathbf{\hat{N}}} = \frac{1}{|\mathbf{\hat{N}}|}  \|\mathbf{\hat{N}} - \mathbf{N}\|_1.}
\end{equation}
Note that the smoothness of the normal map has a significant impact on system performance. Therefore, we introduce an additional smoothing term to encourage a smooth normal distribution among neighboring pixels:
\begin{equation}
    \mathcal{L}_{\text{smooth}} = \sum_{m,n} \left( \left| \nabla_m \mathbf{\hat{N}}_{m,n} \right| + \left| \nabla_n \mathbf{\hat{N}}_{m,n} \right| \right),
\end{equation}
where $\mathbf{\hat{N}}_{m,n}$ represents the estimated normal vector at the $(m, n)$-th pixel and $\nabla$ represents the finite difference operator.
%that convolves its input with $[-1, 1]$ for the $m$-axis and $[-1, 1]^\top$ for the $n$-axis. 
To sum up, the final normal regularization loss is the summation of the $\ell_1$-loss and the smoothing loss:
\begin{equation}
   {\mathcal{L}_{\text{normal}} = \mathcal{L}_{\mathbf{\hat{N}}} + \mathcal{L}_{\text{smooth}}.} 
\end{equation}

\subsubsection{Gaussian Initialization}\label{SubSec_GI}
As illustrated in \cite{kerbl3Dgaussians}, initialization methods for the 3D Gaussian primitives not only affect the final rendering quality but also determine whether the model can converge. For this purpose, we use the depth maps generated by monocular models  illustrated in \eqref{eq:aligned_depth_map} to initialize the Gaussian primitives, which enables the model to learn more accurate geometric details compared to sparse SfM-based depth.

\begin{figure*}
    \centering
    \includegraphics[width=0.95\linewidth]{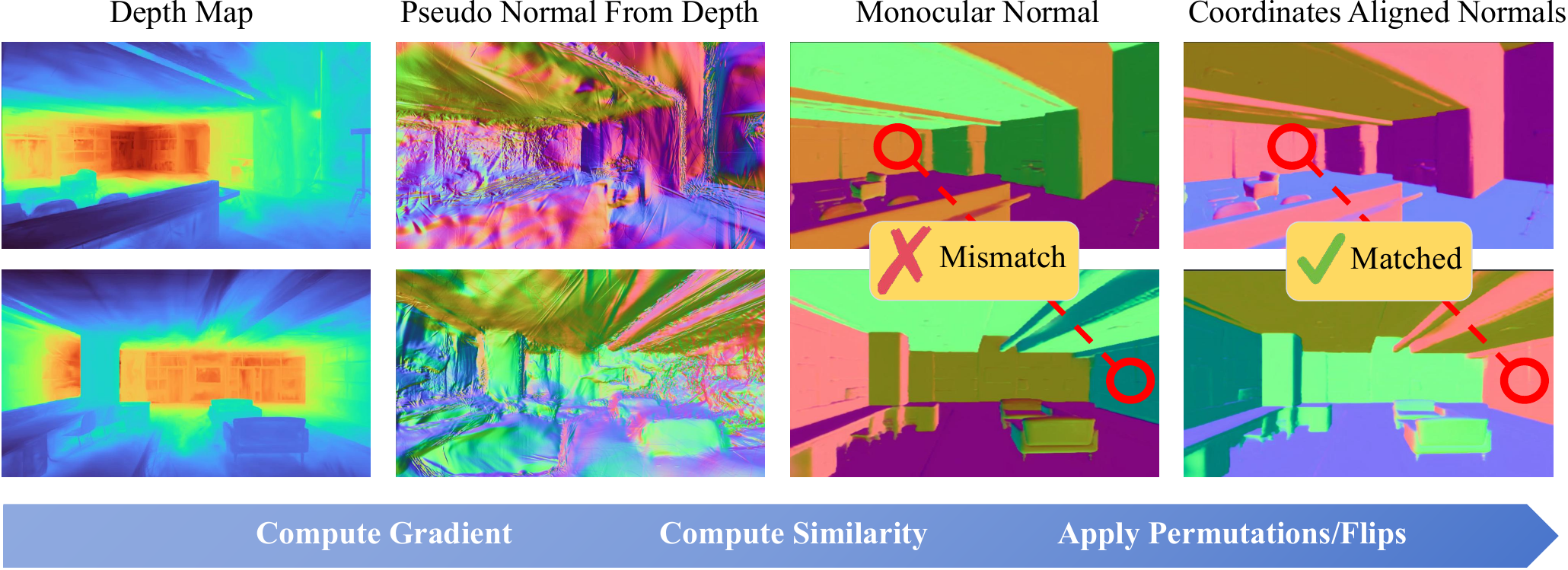}
    \caption{\textbf{Coordinate alignment of monocular normal maps using depth-derived pseudo normals.}}
    \label{fig:normal}
\end{figure*}

For normal initialization, we use depth gradients to calculate pseudo normals \cite{chen2024gi}. While this may introduce some irregularities due to depth noise, it ensures consistent coordinate system alignment and maintains accurate global directional coherence within the field of view.
Based on these observations, we introduce a pre-processing scheme for
normal coordinate alignment. To start with, coarse estimates of the normal vectors are obtained by computing the gradients of the depth map. Second, we enumerate all plausible coordinate transformations for the monocular normal map, including channel permutations and directional flips, and then identify the transformation that yields the highest cosine similarity to the coarse estimate. This approach effectively resolves the coordinate system inconsistencies inherent in monocular normal maps. The detailed pre-processing scheme is illustrated in Fig.~\ref{fig:normal}. 

In summary,
noticing the fact that the precise depth and normal information of the scene are critical to model the propagation of electromagnetic waves,
we utilize these information obtained from visual devices as geometric constraints during the training of the proposed URF-GS model. 
As shown in the subsequent subsection, the URF-GS model trained with geometric constraints provides a robust foundation for accurate channel modeling in the wireless domain.

\subsection{Wireless Domain}\label{Sec_WD}
We model the wireless radiation field using the same set of the 3D-GS primitives optimized for the optical domain in Section \ref{Sec_OD}. The goal is to predict the spatial power distribution with arbitrary Tx-Rx locations. To achieve this, rather than performing direct rasterization for wireless modeling in vanilla 3D-GS literature \cite{wen2025wrf, zhang2024rf}, URF-GS adopts the physical principles to simulate the radio signal propagation. In particular, we decompose the radio signal propagation into the \emph{surface interaction} part and the \emph{free-space transmission} part. This enables more accurate and rapid spatial spectrum prediction under arbitrary Tx-Rx configurations.

In this subsection, we first introduce the surface interaction pattern in Section~\ref{SubSec_PBR} based on the physically-based rendering (PBR). Then, we formulate the free-space transmission process in Section~\ref{SubSec_WRF}. Finally, we conclude this section and summarize the entire URF-GS framework in Section~\ref{SubSec_CURF}.

\subsubsection{Surface Interaction Pattern}\label{SubSec_PBR}
Recovering geometry, materials, and signals from a scene has been a long-standing challenge due to the complex interactions between signals and objects. PBR is a rendering technique that models the appearance of a surface based on the interaction between visible light and its material properties. With the energy conservation principle, PBR is capable of producing results that are consistent with the real world measurements. To avoid solving complex Maxwell's equations and further improve the universality of the model, we simplify the propagation of wireless EM waves in space following PBR. Given a surface point \( \mathbf{x} \) and the corresponding normal vector \( \mathbf{n} \), the outgoing signal $S_{o}$ can be calculated using the following equation \cite{kajiya1986rendering}:
\begin{equation}
    S_o \left( \mathbf{\omega_o}, \mathbf{x} \right) = \int_{\Omega} f_{\text{brdf}} \left( \mathbf{\omega_o}, \mathbf{\omega_i}, \mathbf{x} \right) S_i \left( \mathbf{\omega_i}, \mathbf{x} \right) \left( \mathbf{\omega_i} \cdot \mathbf{n} \right) d\mathbf{\omega_i},
\end{equation}
where  \( \Omega \) is the upper hemisphere centered at \( \mathbf{x} \) and \( f_{\text{brdf}} \left( \mathbf{\omega_o}, \mathbf{\omega_i}, \mathbf{x} \right) \) is the bidirectional reflectance distribution function (BRDF) that characterizes the relationship between the outgoing signal \( S_o \) and the incident one \( S_i \) coming from direction  \( \mathbf{\omega_i} \) based on the material of the object. In the wireless domain, the BRDF can be divided into a scattering component \( f_s \) and a reflection component \( f_r \), according to the Cook-Torrance BRDF model \cite{cook1982reflectance} as follows:
\begin{equation}
     f_{\text{brdf}}(\mathbf{\omega}_i, \mathbf{\omega}_o) = f_s + f_r(\mathbf{\omega}_i, \mathbf{\omega}_o).
     \label{eq:brdf}
\end{equation}
The scattering component $f_s$ can be further expressed as 
\begin{equation}
    f_s = (1 - m) {\mathbf{a}}/{\pi},
    \label{eq:fs}
\end{equation}
where \( \mathbf{a} \in [0, 1]^3 \) denotes the scattering albedo that characterizes the proportion of signals scattered in the 3D space and \( m \in [0, 1] \) represents the metallic value, which indicates the sensitivity to polarization and frequency.

The reflection component \( f_r \) is represented as:
\begin{equation}
    f_r(\mathbf{\omega}_i, \mathbf{\omega}_o) = \frac{D(\mathbf{h;\rho}) F(\mathbf{\omega}_o,\mathbf{h};\mathbf{a},m) G(\mathbf{\omega}_i,\mathbf{\omega_o},\mathbf{h};\rho)}{4 (\mathbf{n} \cdot \mathbf{\omega}_i) (\mathbf{n} \cdot \mathbf{\omega}_o)},
    \label{eq:fr}
\end{equation}
where \( \rho \in [0, 1] \) denotes the roughness of the material, which reflects the flatness of the surface normals and determines the directionality of electromagnetic wave reflection, and \( \mathbf{h} \triangleq ({\mathbf{\omega}_i + \mathbf{\omega}_o})/{\|\mathbf{\omega}_i + \mathbf{\omega}_o\|} \) is used to simplify the calculation.
As shown in \eqref{eq:fr}, the reflection component \( f_r(\mathbf{\omega}_i, \mathbf{\omega}_o) \) can be decomposed into three parts, the normal distribution function \( D \), the Fresnel term \( F \), and the geometry term \( G \). In particular, \( D \) describes the concentration of reflected energy around the specular direction via microscale normal statistics, \( F \) captures angle-dependent reflectivity tied to medium dielectric constants, and \( G \) accounts for microstructural shading effects on effective reflection. 

Based on \eqref{eq:brdf}, the outgoing signal \( S_o \) can also be decomposed into a scattering component \( S_s \) and a reflection component \( S_r \) as follows:
\begin{equation}
S_o \left( \mathbf{\omega}_o, \mathbf{x} \right) = S_s \left( \mathbf{x} \right) + S_r \left( \mathbf{\omega}_o, \mathbf{x} \right),
\end{equation}
where $S_s \left( \mathbf{x} \right) = \int_{\Omega} f_s S_i \left( \mathbf{\omega}_i, \mathbf{x} \right) \left( \mathbf{\omega}_i \cdot \mathbf{n} \right) d\mathbf{\omega}_i$ and
\begin{equation}
 S_r \left( \mathbf{\omega}_o, \mathbf{x} \right) = \int_{\Omega} f_r \left( \mathbf{\omega}_i, \mathbf{\omega}_o \right) S_i \left( \mathbf{\omega}_i, \mathbf{x} \right) \left( \mathbf{\omega}_i \cdot \mathbf{n} \right) d\mathbf{\omega}_i.
\end{equation}
We further define the attenuation coefficient $A(\mathbf{\omega_o}, \mathbf{\omega_i}, \mathbf{x}) \triangleq {S_o \left( \mathbf{\omega}_o, \mathbf{x} \right)}/{S_i \left( \mathbf{\omega_i, \mathbf{x}}\right)}$, which calculates the ratio of the outgoing signal from $\mathbf{\omega}_o$ to the incoming signal from $\mathbf{\omega}_i$ at position $\mathbf{x}$. 

\subsubsection{Free-Space Transmission Pattern}\label{SubSec_WRF}
\begin{figure}
    \centering
    \includegraphics[width=0.6\linewidth]{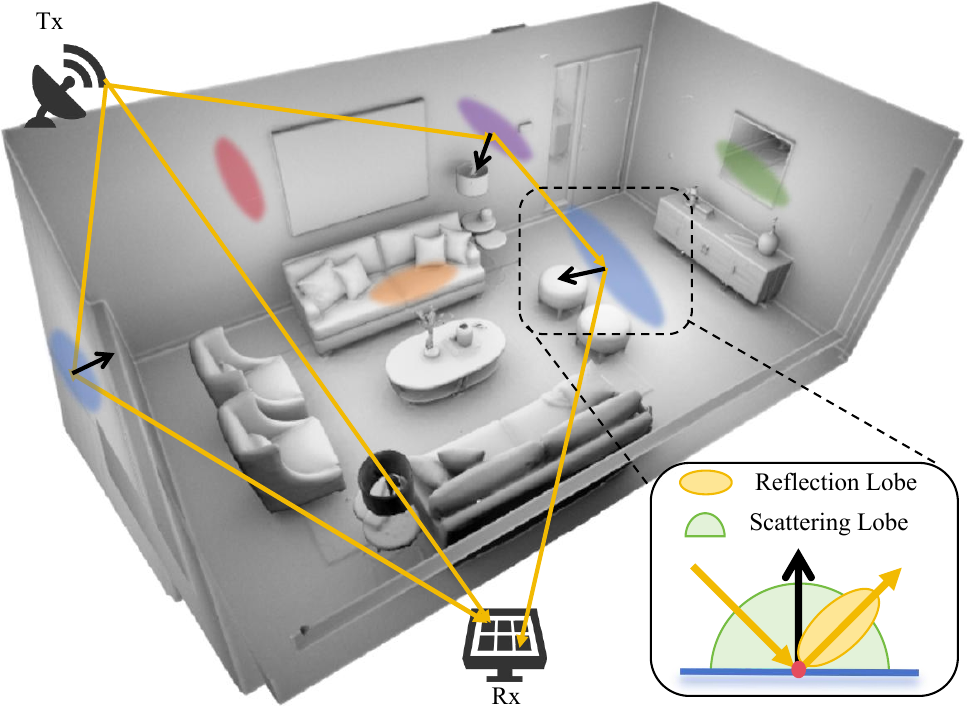}
    \caption{\textbf{An illustration of physically-based inverse rendering for the radio signal propagation modeling.}}
    \label{fig:fullpath}
\end{figure}

Unlike optical rendering in vanilla 3DGS-based PBR, which typically neglects light attenuation over distance, wireless signals experience significant propagation loss. Therefore, we incorporate the free space path loss (FSPL) model into our framework to capture this effect. The signal propagation modeling process for the entire scene is shown in Fig. \ref{fig:fullpath}. The Tx with transmit power $P_\text{Tx}$ emits a ray whose gain equals to \( G_{\text{Tx}}(\theta_{{d}}, \phi_{{d}}) \), where \( (\theta_{\text{d}}, \phi_{\text{d}}) \) corresponds to the angle of departure (AoD). The ray reaches the Rx after \( L \) interactions with object surfaces, with a gain of \( G_{\text{Rx}}(\theta_{{a}}, \phi_{{a}}) \), where $(\theta_{{a}}, \phi_{{a}})$ corresponds to its angle of arrival (AoA). The power \( P_{Rx} \) at the receiver can be expressed as:
\begin{equation}
    P_{\text{Rx}} = P_{\text{Tx}} G_{\text{Tx}}(\theta_{{d}}, \phi_{{d}})  G_{\text{Rx}}(\theta_{{a}}, \phi_{{a}}) \text{FSPL} {\prod_{\ell=1}^L A_{\ell}},
\end{equation}
with
\begin{equation}
    % \text{FSPL} =  {\left(\frac{c}{4\pi f (\sum_{\ell=1}^{L+1} d_{\ell})}\right)^2} = \textcolor{red}{\prod_{\ell=1}^{L+1} \left(\frac{c}{4\pi f d_\ell}\right)^2},
    \text{FSPL} =  \prod_{\ell=1}^{L+1} \left(\frac{c}{4\pi f d_\ell}\right)^2,
\end{equation}
where $c$ represents the speed of light, $f$ is the carrier frequency, and $d_\ell$ corresponds to the path length between the $(\ell - 1)$-th and the $\ell$-th interactions.

This multiplicative form is adopted to explicitly capture the cascaded attenuation of the signal along each propagation segment, a process that is physically equivalent to applying the inverse-square law sequentially. In the limit where diffuse reflections dominate the multipath structure, this formulation aligns naturally with the radar equation principle, where the total path loss factorizes into contributions from each segment of the propagation chain. While mathematically equivalent to evaluating the FSPL directly from the total path length, the per-segment product representation provides a more interpretable decomposition of the end-to-end attenuation and integrates seamlessly with the per-interaction surface response coefficients $A_\ell$.
%\textcolor{blue}{Since we consider $L$ interactions with the objects in the scene, the number of paths for FSPL is $L+1$.}

Based on the above analysis, in addition to the geometry attributes in Section \ref{Sec_OD}, we augment each 3D Gaussian primitive with material attributes. Specifically, we introduce albedo, metallic, roughness, and normal attributes to each 3D Gaussian primitive, which correspond to the variables $\mathbf{a}, m$ and $\rho$, respectively, that are introduced in \eqref{eq:fs} and \eqref{eq:fr}.
%By integrating FSPL and surface scattering into the Tx-Rx-environment interaction loop, our simulation strictly follows electromagnetic propagation laws.
Importantly, the proposed URF-GS framework is fully differentiable and the learnable attributes of the Gaussian primitives are optimized via gradient descent.

\subsubsection{The Constructed URF and Radio Map}\label{SubSec_CURF}

In summary, URF-GS integrates optical reconstruction and wireless propagation through a two-stage pipeline based on differentiable 3D Gaussian primitives that are shared by two tasks. In the first stage, pre-trained monocular networks are adopted to estimate the depth and normal, which guide the optimization of 3D Gaussians to obtain accurate 3D geometry. In the second stage, a physics-aware inverse rendering scheme is employed to explicitly model scattering, reflection, and attenuation of the wireless signal to jointly refine material properties and radiation patterns. In the inference phase, the trained URF-GS model generates both optical images and radio maps for different Tx–Rx configurations without requiring additional training, thereby producing accurate and generalizable 3D radio maps.

\subsection{Implementation Details}\label{Sec_IDs}

URF-GS is implemented in PyTorch and evaluated on a server equipped with an Intel Xeon Gold 6230R CPU and an NVIDIA GeForce RTX 3090 GPU. The key implementation strategies and configurations are as follows:

\textbf{Data Caching.} The optical pipeline requires concurrent access to a large volume of RGB images, depth maps, and normal maps. Storing all predicted maps simultaneously can exceed the capacity of GPUs with limited VRAM (e.g., less than 24 GB). To mitigate this, we implement a dynamic loading strategy with a threshold of 500 samples. When the dataset size exceeds this limit, the excess data is cached in the CPU memory. Tensors are transferred to the GPU on demand and released immediately after processing, ensuring efficient memory management.

\textbf{Deferred Shading.} To accurately compute multi-bounce signal interactions within the 3D environment, we adopt a deferred shading strategy~\cite{chen2024gi}. This approach first rasterizes normal, depth, and BRDF maps into a G-buffer. Then, a differentiable path tracer is used to query these geometric attributes to simulate signal–environment interactions, including complex propagation paths and occlusions.

\textbf{Optimization Details.} URF-GS utilizes a unified representation for both optical and wireless reconstruction. For the optical component, we follow the standard 3D-GS framework, parameterizing colors via spherical harmonics for $\alpha$-blending and employing adaptive density control to balance geometric fidelity with rendering efficiency. For radio signal propagation, we integrate PBR attributes, i.e., specifically albedo, roughness, and metallic parameters, into the Gaussians. This ensures consistent material modeling across both radio and optical domains. All learnable parameters are jointly optimized using the Adam optimizer~\cite{adam2014method}, consisting of $3\times10^4$ iterations for optical reconstruction followed by $10^4$ iterations for wireless reconstruction to ensure convergence.

\section{Data Availability}\label{sec5}
The datasets generated and analyzed during the current study are available from the corresponding author upon reasonable request, and will be deposited in a public repository upon acceptance of the manuscript.

\section{Code Availability}\label{secCode}
All relevant code supporting the results of this study will be made publicly available in a dedicated open-source repository upon acceptance of the manuscript.

\section{Acknowledgments}\label{sec6}
This work was supported by the Hong Kong Research Grants Council under the Areas of Excellence scheme grant AoE/E-601/22-R and NSFC/RGC Collaborative Research Scheme grant CRS\_HKUST603/22.

\section{Author Contributions}
C.W. and J.T. conceived the study and designed the methodology. C.W. and J.T. implemented the software and conducted the experiments. C.W., C.B., and Z.L. analyzed the results. J.Z. supervised the project and acquired funding. All authors contributed to writing and reviewing the manuscript.

\section{Competing Interests}
The authors declare no competing interests.

\bibliographystyle{plainnat}
\bibliography{paper}

\end{document}